\documentclass[5p,authoryear,preprint,12pt]{elsarticle}
\makeatletter\if@twocolumn\PassOptionsToPackage{switch}{lineno}\else\fi\makeatother

\usepackage{tabulary,xcolor}
\usepackage{amsfonts,amsmath,amssymb}
\usepackage[T1]{fontenc}
\makeatletter
\let\save@ps@pprintTitle\ps@pprintTitle
\def\ps@pprintTitle{\save@ps@pprintTitle\gdef\@oddfoot{\footnotesize\itshape \null\hfill\today}}
\def\hlinewd#1{%
  \noalign{\ifnum0=`}\fi\hrule \@height #1%
  \futurelet\reserved@a\@xhline}

\def\tblbottomrule{\hlinewd{.8pt}}
\def\tblmidrule{\noalign{\vspace*{6pt}}\hline\noalign{\vspace*{2pt}}}
\AtBeginDocument{\ifNAT@numbers \biboptions{sort&compress}\fi}
\makeatother

\usepackage{ifluatex}
\ifluatex
\usepackage{fontspec}
\defaultfontfeatures{Ligatures=TeX}
\usepackage[]{unicode-math}
\unimathsetup{math-style=TeX}
\else 
\usepackage[utf8]{inputenc}
\fi 
\ifluatex\else\usepackage{stmaryrd}\fi

\usepackage{url,multirow,morefloats,floatflt,cancel,tfrupee}
\makeatletter

\AtBeginDocument{\@ifpackageloaded{textcomp}{}{\usepackage{textcomp}}}
\makeatother
\usepackage{colortbl}
\usepackage{xcolor}
\usepackage{pifont}
\usepackage[nointegrals]{wasysym}
\makeatletter

\def\mcWidth#1{\csname TY@F#1\endcsname+\tabcolsep}

\def\cAlignHack{\rightskip\@flushglue\leftskip\@flushglue\parindent\z@\parfillskip\z@skip}
\def\rAlignHack{\rightskip\z@skip\leftskip\@flushglue \parindent\z@\parfillskip\z@skip}

\usepackage{ifxetex}
\ifxetex\else\if@twocolumn\usepackage{dblfloatfix}\fi\fi

\AtBeginDocument{
\expandafter\ifx\csname eqalign\endcsname\relax
\def\eqalign#1{\null\vcenter{\def\\{\cr}\openup\jot\m@th
  \ialign{\strut$\displaystyle{##}$\hfil&$\displaystyle{{}##}$\hfil
      \crcr#1\crcr}}\,}
\fi
}

\AtBeginDocument{%
  \@ifpackageloaded{endfloat}%
   {\renewcommand\efloat@iwrite[1]{\immediate\expandafter\protected@write\csname efloat@post#1\endcsname{}}}{}%
}%

\def\BreakURLText#1{\@tfor\brk@tempa:=#1\do{\brk@tempa\hskip0pt}}
\let\lt=<
\let\gt=>
\def\processVert{\ifmmode|\else\textbar\fi}

\@ifundefined{subparagraph}{
\def\subparagraph{\@startsection{paragraph}{5}{2\parindent}{0ex plus 0.1ex minus 0.1ex}%
{0ex}{\normalfont\small\itshape}}%
}{}

\newcommand\role[1]{\unskip}
\newcommand\aucollab[1]{\unskip}
  
\@ifundefined{tsGraphicsScaleX}{\gdef\tsGraphicsScaleX{1}}{}
\@ifundefined{tsGraphicsScaleY}{\gdef\tsGraphicsScaleY{.9}}{}
\def\checkGraphicsWidth{\ifdim\Gin@nat@width>\linewidth
	\tsGraphicsScaleX\linewidth\else\Gin@nat@width\fi}

\def\checkGraphicsHeight{\ifdim\Gin@nat@height>.9\textheight
	\tsGraphicsScaleY\textheight\else\Gin@nat@height\fi}

\def\fixFloatSize#1{}%\@ifundefined{processdelayedfloats}{\setbox0=\hbox{\includegraphics{#1}}\ifnum\wd0<\columnwidth\relax\renewenvironment{figure*}{\begin{figure}}{\end{figure}}\fi}{}}
\let\ts@includegraphics\includegraphics

\def\inlinegraphic[#1]#2{{\edef\@tempa{#1}\edef\baseline@shift{\ifx\@tempa\@empty0\else#1\fi}\edef\tempZ{\the\numexpr(\numexpr(\baseline@shift*\f@size/100))}\protect\raisebox{\tempZ pt}{\ts@includegraphics{#2}}}}

\AtBeginDocument{\def\includegraphics{\@ifnextchar[{\ts@includegraphics}{\ts@includegraphics[width=\checkGraphicsWidth,height=\checkGraphicsHeight,keepaspectratio]}}}

\DeclareMathAlphabet{\mathpzc}{OT1}{pzc}{m}{it}

\def\URL#1#2{\@ifundefined{href}{#2}{\href{#1}{#2}}}

\def\UrlOrds{\do\*\do\-\do\~\do\'\do\"\do\-}%
\g@addto@macro{\UrlBreaks}{\UrlOrds}

\@ifundefined{quoteAttrib}
	{}
	{}

\@ifundefined{titlequoteAttrib}
	{}{}

\newenvironment{title-quote}
	{\list{}{\fontsize{10pt}{12pt}\selectfont\leftmargin.5in\itshape\rightmargin\leftmargin}%
  \item\relax}
  {\endlist}

\makeatother

\makeatletter
\def\ps@pprintTitle{\save@ps@pprintTitle\gdef\@oddfoot{\footnotesize\hfill\thepage\itshape \null\hfill\today}}
\makeatother
          
\usepackage{float}

\begin{document}

\begin{frontmatter}
	
\title{The Wireless Control Plane: An Overview and Directions for Future Research
}
    
\author[aff6ef6c6d3ca1f4250cbdee53b5d68a6ec,aff54d782fe43b83cf8f26685ef9c0bb9d5,aff980730d7a3b112f78fe168facbc477f8,aff7c71628829ff8298b720136708dc69fd]{EmadelDin A. Mazied\corref{contrib-e917233ea06b9af50996a4e90c801db9}}
\ead{emazied@eng.sohag.edu.eg}\cortext[contrib-e917233ea06b9af50996a4e90c801db9]{Corresponding author.}
\author[aff980730d7a3b112f78fe168facbc477f8,affb1fba3c25ea8077d1715926c98391460]{Mustafa Y. ElNainay}
\ead{ymustafa@alexu.edu.eg}
\author[aff980730d7a3b112f78fe168facbc477f8,aff657827ba591f2d520ce4b1df6189dbfa]{Mohammad J. Abdel-Rahman}
\ead{mo7ammad@vt.edu}
\author[aff980730d7a3b112f78fe168facbc477f8]{Scott F. Midkiff}
\ead{midkiff@vt.edu}
\author[aff54d782fe43b83cf8f26685ef9c0bb9d5]{Mohamed R. M. Rizk}
\ead{mohamed.rizk@alexu.edu.eg}
\author[aff7c71628829ff8298b720136708dc69fd]{Hesham A. Rakha}
\ead{hrakha@vtti.vt.edu}
\author[aff980730d7a3b112f78fe168facbc477f8]{Allen B. MacKenzie}
\ead{mackemab@vt.edu}
    
\address[aff6ef6c6d3ca1f4250cbdee53b5d68a6ec]{Electrical Engineering\unskip, 
    Sohag University\unskip, New Sohag\unskip, 82511\unskip, Sohag\unskip, EGYPT}
  	
\address[aff54d782fe43b83cf8f26685ef9c0bb9d5]{Electrical Engineering\unskip, 
    Alexandria University\unskip, 21544\unskip, Alexandria\unskip, EGYPT}
  	
\address[aff980730d7a3b112f78fe168facbc477f8]{The Bradley Department of Electrical and Computer Engineering\unskip, 
    Virginia Tech\unskip, Blacksburg\unskip, 24061\unskip, VA\unskip, USA}
  	
\address[aff7c71628829ff8298b720136708dc69fd]{Virginia Tech Transportation Institute\unskip, 
    Center for Sustainable Mobility\unskip, Virginia Tech\unskip, Blacksburg\unskip, 24061\unskip, VA\unskip, USA}
  	
\address[affb1fba3c25ea8077d1715926c98391460]{Computers and Systems Engineering\unskip, 
    Alexandria University\unskip, 21544\unskip, Alexandria\unskip, EGYPT}
  	
\address[aff657827ba591f2d520ce4b1df6189dbfa]{Electrical Engineering\unskip, 
    Al Hussein Technical University (HTU)\unskip, 11821\unskip, Amman\unskip,  Jordan}

\begin{abstract}
Software-defined networking (SDN), which has been successfully deployed in the management of complex data centers, has recently been incorporated into a myriad of 5G networks to intelligently manage a wide range of heterogeneous wireless devices, software systems, and wireless access technologies. Thus, the SDN control plane needs to communicate wirelessly with the wireless data plane either directly or indirectly. The uncertainties in the wireless SDN\mbox{}\protect\newline control plane (WCP) make its design challenging. Both WCP schemes (direct WCP, D-WCP, and indirect WCP, I-WCP) have been incorporated into recent 5G networks; however, a discussion of their design principles and their design limitations is missing. This paper introduces an overview of the WCP design (I-WCP and D-WCP) and discusses its intricacies by reviewing its deployment in recent 5G networks. Furthermore, to facilitate synthesizing a robust WCP, this\mbox{}\protect\newline paper proposes a generic WCP framework using deep reinforcement learning (DRL) principles and presents a roadmap for future research.
\end{abstract}
  
\end{frontmatter}
    
\section{Introduction}
The next generation wireless network (5G) is envisioned to benefit humanity by offering a quality of life instead of only achieving a quality of service. In this regard, 5G has been proposed to provide a communication infrastructure to Internet-of-Things (IoT) systems for human health and comfort (e.g., IoT systems contribute to managing smart homes, supporting health-care applications, providing healthy foods by using smart agriculture techniques, and detecting pollution in the surrounding environment) \unskip~\citet{314005:7022254,314005:7022255}. Moreover, 5G has recently been deployed in Intelligent Transportation System (ITS) applications for human safety \unskip~\citet{314005:7022851}. 

To tackle these sophisticated design requirements for 5G networks, all layers of the 5G protocol stack need to be rapidly developed in the presence of an agile and robust management framework. This requires incorporating flexible and rapidly evolving software-based solutions into 5G systems rather than relying on inflexible and slowly evolving hardware-based design approaches \unskip~\citet{314005:7022254}.

\bgroup
\fixFloatSize{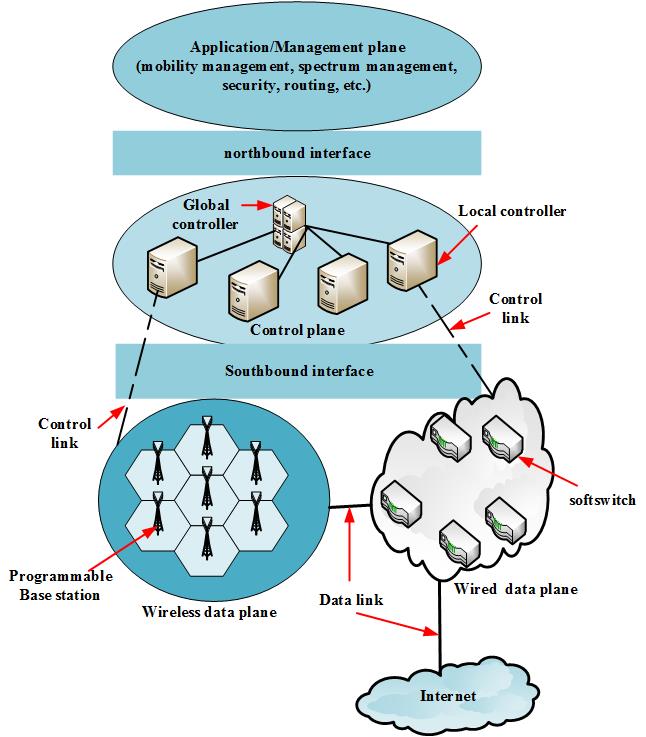}
\begin{figure*}[!htbp]
\centering \makeatletter\IfFileExists{74fd87f9-354e-4b31-9a39-1e8539b4b40d-ufig01_sdnarchit.jpg}{\includegraphics{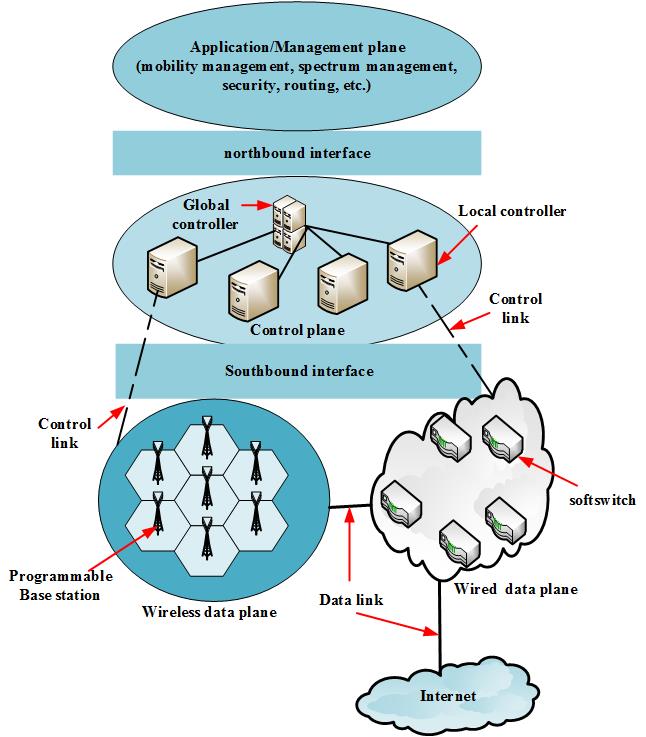}}{}
\makeatother 
\caption{{The top-down SDN architecture.}}
\label{figure-28ab42b389039cdaa87e60d9e2b8999d}
\end{figure*}
\egroup

The software-defined network (SDN) is complemented with a software-defined radio (SDR) to bring a holistic software-based solution for the 5G network. While the softwarization of the physical (PHY) and medium access control (MAC) layers of the protocol stack is implemented through SDR methods, the softwarization of higher layers of the protocol stack, along with providing a programmable management framework for a networked system, is achieved using SDN principles \unskip~\citet{314005:7022254,314005:7022256,314005:7022257,314005:7022258,314005:7022260,314005:7022261}. SDN abstracts the network's control functions by decoupling them from their associated data forwarding infrastructure and implements them in a software that runs on a central server (SDN controller), which reasons for the whole network. Figure~\ref{figure-28ab42b389039cdaa87e60d9e2b8999d} depicts the SDN architecture, which comprises the data plane (DP), control plane (CP), and management plane (MP), where the CP is key to the SDN concept.

Recent proposals for 5G networks have considered SDN for their management. For instance, \unskip~\citet{314005:7022258} utilized the global view of the centralized SDN controller for ensuring end-to-end (ETE) connectivity for unmanned aerial vehicles (UAVs) as well as overcoming the limitations of route determination, frequent link failures, and UAV's limited onboard processing resources. In this system, the CP collects the necessary information from the UAV DP to implement multi-path routing among UAVs to attain an elastic ETE connection scheme.

As the CP needs to communicate with a diverse range of heterogeneous wireless DP (WDP) elements, different 5G network proposals have deployed SDN in which the CP communicates with the WDP using wireless connections\unskip~\citet{314005:7022256,314005:7022258,314005:7022260,314005:7022261}. \unskip~\citet{314005:7022258} proposed CP that communicated with the UAV DP using wireless connections because the use of wired connections between the SDN controller and mobile UAVs is not practically feasible\unskip~\citet{314005:7022258,314005:7022259}. In this sense, there are two wireless CP\footnote{Throughout the paper, wireless CP refers to the SDN-based 5G network in which the CP communicates with the underlying WDP over wireless channels either directly or indirectly.}(WCP) classes, i.e., direct\unskip~\citet{314005:7022256} and indirect\unskip~\citet{314005:7022495}. In the former, direct wireless connections are used between the controller and the underlying WDP. In the latter, the controller communicates with the underlying WDP through a master wireless DP, such as a roadside unit (RSU), which communicates with the other WDP elements (vehicles) that are in the vicinity of it using wireless connections. 

Although the WCP has been widely suggested in several 5G networks, the discussion of its design principles is superficially examined. In the literature, the discussion of the WCP design was presented from the controller placement perspective\unskip~\citet{314005:7022498,314005:7022499,314005:7022406,314005:7022253}. In this sense, the controllers are optimally distributed within a wireless environment to satisfy a specific design requirement (e.g., scalability\unskip~\citet{314005:7022498,314005:7022499}\unskip~\citet{314005:7022406} or reliability\unskip~\citet{314005:7022253}) in the presence of a wide range of uncertainties\footnote{Wireless environment introduces different levels of uncertainties, such as uncertainty in base station loads, uncertainty in retransmissions, uncertainty in interference levels, and uncertainty in malicious activities.}. Nevertheless, identifying the strengths and the weaknesses of each WCP scheme and discussing their design challenges has been overlooked. Therefore, rethinking the WCP design to pave the way for synthesizing a standardized WCP for 5G networks and beyond becomes urgently needed since both WCP categories introduce the major challenge of transmitting critical control information within a wireless environment. 

This paper takes a step in the direction of discussing the WCP design principles for the 5G networks and beyond. In particular, this paper introduces:

\begin{itemize}
  \item \relax An Overview of the WCP deployment in different 5G networks,
  \item \relax A Qualitative comparison between the deployment of an indirect WCP and a direct WCP in the recent proposals for 5G networks,
  \item \relax A generic WCP framework that could tackle the diverse range of uncertainties in WCP environment.
\end{itemize}
  Inspired by the successful deployment of deep reinforcement learning (DRL) in several sophisticated control systems that are replete with uncertainties (e.g., robotics)\unskip~\citet{314005:7022311}, the generic WCP framework is proposed by using DRL tool. In this sense, DRL could address the WCP's design challenges since it has an agent (SDN controller) that makes its control decision (control action) according to its interaction with an environment (SDN controlled system). In fact, there is a mutual benefit between SDN and DRL. As the SDN controller maintains the status of all the network's entities, it benefits the DRL by providing it with a pool of data sources that interpret the dynamics of the environment states. Furthermore, DRL benefits the SDN through its flexibility to be implemented by different higher level languages (e.g., C++ and Python), which facilitates its integration with the emerged SDN programming language (Pyretic)\unskip~\citet{314005:7022317}.

The remainder of this paper is organized as follows. A review of the SDN principle and a discussion of its significance for 5G networks are presented in Section 2. Next, a review of the WCP use cases in recent proposals for 5G networks, such as software-defined UAVs (SDUAVs)\unskip~\citet{314005:7022258}, software-defined ultra-dense networks (SDUDNs)\unskip~\citet{314005:7022256,314005:7022265}, software-defined Internet-of-Vehicles (SDIoV)\unskip~\citet{314005:7022260,314005:7022320}, and software-defined IoT (SDIoT)\unskip~\citet{314005:7022261}, along with highlights of the potential benefits of its deployment in each scenario, is presented in Section 3. Following that, in Section 4, a WCP framework is proposed. Subsequently, in Section 5, a few interesting problems for future research are underlined. Finally, the paper is concluded in Section 6.
    
\section{SDN for 5G and Beyond }
In this section, a brief description of the SDN principles along with a discussion on its benefits for 5G networks is presented. Moreover, the SDN control plane design principles are reviewed.

\subsection{What is SDN?}SDN is primarily developed based on modular design, network programmability\footnote{Programmability refers to the tendency of networks to be programmable, which lies in the fact that the network structure resembles the program structure in abstraction and recursive, that layering of the network protocol abstracts many processes, and that Internet hierarchy is recursive because it is an inter-network of small-scale networks in which each small-scale network comprises a set of smaller base networks.}, and centralized control. Accordingly, SDN defeats the network complexity by decomposing the network's control functions, decoupling them from their associated forwarding devices and abstracting them in a centralized and programmable control entity (i.e., the SDN controller). In this sense, the SDN architecture constitutes three planes, as shown in Figure~\ref{figure-28ab42b389039cdaa87e60d9e2b8999d}: the data plane (DP), which represents the forwarding data device infrastructure; the control plane (CP, which comprises the network logic to reason for the whole network; and the management plane (MP), which provides a software development environment for network developers to develop innovative management solutions (e.g., admission control, traffic control and routing policy)\unskip~\citet{314005:7023628}.

\subsection{SDN for 5G networks and beyond}5G anticipates offering new and rapidly evolving services to connect the ever-increasing smart devices, along with providing ultra-reliable and ultra-low latency end-to-end (ETE) services for critical 5G applications. In this regard, there is an urgent need to not only rethink the design of all layers of the protocol stack but also develop new cross-layer design techniques because they have offered performance improvements in a myriad of critical network applications, for instance, optimizing the throughput and delay performance of wireless sensor networks (WSNs) that support health and environmental applications\unskip~\citet{314005:7022266}. Fortunately, SDN contributes to achieving the 5G vision with its centralized and programmable control entity, where all required cross-layer information is available within it. In fact, SDN fits all network pieces together, in the wired and wireless domains, to deliver innovative network solutions for improved ETE performance for 5G applications. \mbox{}\protect\newline In\unskip~\citet{314005:7022312}, for example, the authors proposed an SDN-based programmable cross-layer optimization framework to ensure the fulfillment of the ETE latency requirement for safety applications in 5G-based ITS, where autonomous vehicles communicate over millimeter-wave (mmWave) channels. Because the mmWave suffers from a limited communication range, high sensitivity to physical obstacles and severe degradation due to climate variations, a smart coordination, and management scheme for mmWave stations has been proposed. According to\unskip~\citet{314005:7022312}, the medium access control (MAC) and transport control (TCP) delays have been jointly minimized by developing an SDN-based cross-layer optimization framework that utilizes the cross-layer information at its centralized entity to achieve the ultra-low ETE latency design requirement for ITS safety application. \mbox{}\protect\newline According to this proposed framework, the SDN control entity not only contributed to providing the MAC and TCP cross-layer information but also utilized, by exploiting its programmable feature, to optimize the network performance in a reactive manner (i.e., mmWave stations have been adaptively coordinated to subside the effects of uncertainties in mmWave channels' states). In other words, the SDN global view of the whole network and its programmability features have been used for minimizing the delay that has been constrained by dynamic variations in the status of mmWave channels.

Accordingly, the major benefits of SDN deployment in 5G networks are:

\begin{itemize}
  \item \relax Fast development: unlike hardware-based solutions that take years to adopt a new technology, the SDN programmability feature offers a rapid-development cycle that satisfies the rapidly evolving applications and flexible architecture design requirements for 5G systems in addition to contributing to the virtualization of wireless resources\unskip~\citet{314005:7022318}. 
  \item \relax Efficient cross-layer optimization framework: the centralized SDN CP has, in real-time, the important cross-layer information of all layers of the protocol stack that enables the design of a robust online cross-layer optimization, which is ultimately needed to boost the ETE performance for critical 5G apps. 
  \item \relax Taming design complexities: managing 5G systems through the SDN modular design principle 
\end{itemize}
  In this sense, the CP is deemed to be the SDN's brain. We refer to\unskip~\citet{314005:7022254,314005:7022257,314005:7022265,314005:7023628,314005:7022318} for more detail about SDN benefits for 5G and beyond.

\subsection{The CP design principles}The CP utilizes simple management modules (the controller's application programming interfaces, APIs) along with a programmable network operating system (NOS) to manage complex network systems in a centralized manner\unskip~\citet{314005:7023628}. As shown in Figure~\ref{figure-d75eb8a6c9d4db89993d989e394f968b}, the CP has physical and logical topologies. The former describes how many controllers the CP constitutes, whether it has a single controller or multiple distributed controllers, and identifies how the distributed controllers are placed and physically connected to each other. The latter identifies the logical operation of the CP's entities (e.g., managing the controllers' computing and memory resources, NOS, APIs, etc.) as well as manages the logical communication among the controllers within the CP itself and between the CP and the other two planes (DP and MP).

\bgroup
\fixFloatSize{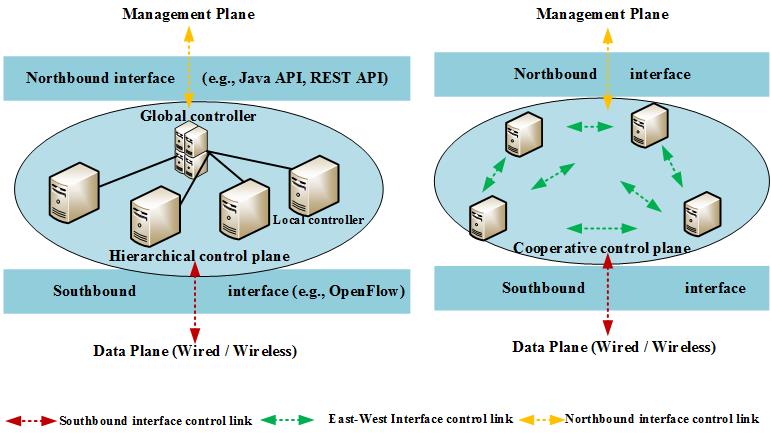}
\begin{figure*}[!htbp]
\centering \makeatletter\IfFileExists{57abc72a-d1fb-45c5-93c0-84673083d534-ufig02_cparchit_updated.jpg}{\includegraphics{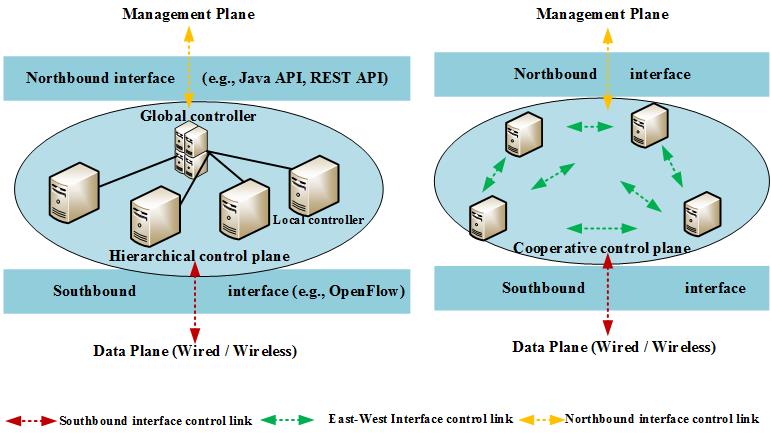}}{}
\makeatother 
\caption{{The SDN Control Plane.}}
\label{figure-d75eb8a6c9d4db89993d989e394f968b}
\end{figure*}
\egroup
Building the CP with either a single controller or multiple controllers has been debated in the literature\unskip~\citet{314005:7022313}. Nevertheless, the deployment of multiple controllers that are physically distributed has widely spread due to the demands for responding to the deluge of DP's requests where DP elements are distributed over large geographical areas\unskip~\citet{314005:7023675,314005:7022263,314005:7022314}. The distributed controllers have to communicate with each other to maintain the SDN's centralized view of the whole network. In this regard, there are two physical topologies:

\begin{itemize}
  \item \relax Hierarchical, in which distributed local controllers are connected to a single global controller to synchronize (centralize) their operation\unskip~\citet{314005:7023675}. 
  \item \relax Cooperative, in which the SDN's global view is maintained through cooperative operation among distributed controllers\unskip~\citet{314005:7023677}. 
\end{itemize}
  Furthermore, there are two logical topologies for the CP with multiple controllers\unskip~\citet{314005:7022314}:

\begin{itemize}
  \item \relax Logically centralized, in which different approaches are deployed to synchronize the operation among distributed controllers, such as Hyperflow, which uses a distributed file system to construct a network's global view. 
  \item \relax Logically distributed, in which each controller is in charge of managing the functions of a specific network domain where the controller-to-controller communication is managed by an east-west interface (EWI) protocol, such as DISCO, which categorizes the controller's functionality as an intra-domain controller or inter-domain one. An intra-domain controller is logically centralized but becomes logically distributed when contributing to an inter-domain operation, such as performing an end-to-end priority service request. 
\end{itemize}
  The second scheme is more popular than the first in large scale distributed networks due to its contribution to attaining a scalable and flexible CP by using an agile EWI protocol\unskip~\citet{314005:7022314}. Most importantly, both logical topologies are workable for either hierarchical or cooperative topologies\unskip~\citet{314005:7023677}.

The CP interacts with the MP through the northbound interface (NI) and interacts with the DP via the southbound interface (SI). The NI manages how a business application talks to a controller to specify its requirements. These requirements are passed down, in the form of programming instructions, to the CP. Consequently, the CP delivers an appropriate response to the MP. This type of CP-MP data exchange is managed by an NI API (e.g., Python API, Java API, Representational State Transfer or REST API)\unskip~\citet{314005:7022318}. Thus, network developers can reconfigure and develop different modular network applications and management policies by using any high level programming language (e.g., Pyretic\unskip~\citet{314005:7022317}) to represent the business requirements. The controller uses the SI to manage the interaction between the CP and the DP. An SI API is utilized to enable a controller to program the underlying DP's elements. OpenFlow is deemed to be the standardized SI API\unskip~\citet{314005:7023628,314005:7022318,314005:7022262,314005:7040583}.
    
\section{The Wireless Control Plane For 5G and Beyond}
The previous section has discussed the significance of SDN deployment in 5G networks and has concluded that the CP is the SDN's brain. In wireless networks, the CP has to communicate with a diverse range of heterogeneous wireless DP (WDP) elements, such as long-term evolution (LTE) base stations (BSs), road side units (RSUs) for vehicular networks, WiFi access points (APs), unmanned aerial vehicles (UAVs) and LTE drones. Thus, deploying a conventional wired CP, where an SDN controller communicates with a DP network device using a wired connection, is not practically possible for a heterogeneous WDP infrastructure. Furthermore, the everlasting increase in WDP elements (i.e., ultra-dense deployment of small-cell base stations, SBSs) limits the flexibility of controller assignment to the WDP elements. Thus, the following subsections present an overview of the wireless CP (WCP), i.e., what the WCP is, what schemes it has, and how it has been deployed in recent 5G systems.

\subsection{The wireless CP }The WCP is the wireless SDN's CP, in which an SDN controller communicates with an WDP element using a wireless connection. Before delving into the discussion of the WCP types for 5G networks, in what follows, a review of the candidate architectures for 5G networks is briefly presented. \mbox{}\protect\newline There are two major candidates: a cloud radio access network (CRAN)\unskip~\citet{314005:7023674} and a cellular network with ultra-dense deployment of SBSs\unskip~\citet{314005:7022256,314005:7022495}. A CRAN comprises a cloud-computing based-band unit (BBU), in which the computing resource pool and virtualization technology are deployed, and radio remote heads (RRHs), which are distributed radio front-ends with antennas that integrate with the BBU to synthesize a CRAN cellular network. The deployed virtualization technology at the BBU cloud enables the creation of different instances of virtual baseband BSs (vBBSs) that process the digitized radio signals, which are received from RRHs. The BBU receives I-Q samples from the RRHs and sends digitized signals to them via a fronthaul network (wireless fronthaul, fiber link-based fronthaul, or hybrid wired and wireless fronthaul)\unskip~\citet{314005:7022265}. In a cellular network with ultra-dense deployment of SBSs (UDN), each cell comprises a single macro-cell BS (MBS) with ultra-dense deployment of SBSs that are distributed within the MBS coverage area. Moreover, SBSs deploy different radio access technologies (RATs) (e.g., LTE, ZigBee, WiFi and radar) to improve the coverage and capacity of the cellular system\unskip~\citet{314005:7022254}. Furthermore, the UDN's umbrella has been extended to include different 5G architectures. For instance, since an SBS could be an AP, RSU, UAV (drone), or a set of some or all previous types that deploy different RATs, the UDN could be seen as a 5G architecture that connects smart vehicles, smart drones, smart sensors, smart homes, and all possible smart wireless things (IoT)\unskip~\citet{314005:7022254,314005:7022261}. In the literature, the WCP was proposed in two major forms.

\textbf{Indirect WCP (I-WCP): }The WDP is partially incorporated into the CP design, in which the CP communicates with the WDP elements through a master WDP element (MWDP). The I-WCP has two tiers of the CP:

\begin{itemize}
  \item \relax The global-tier, consisting of the SDN controllers that communicate with the MWDP (i.e., MBS) via fiber links.
  \item \relax The local-tier, consisting of software-based local agents that reside on the MWDP. 
\end{itemize}
  Through its local agent, the MWDP communicates with its associated WDP elements (i.e., SBSs, RSUs, or drones) using a wireless backhaul\unskip~\citet{314005:7022495}. 

\textbf{Direct WCP (D-WCP): }The controller directly communicates with the WDP elements using wireless connections, e.g.,\unskip~\citet{314005:7022256,314005:7023627}.depicts the two different WCP-to-WDP communication scenarios for the 5G networks.

\bgroup
\fixFloatSize{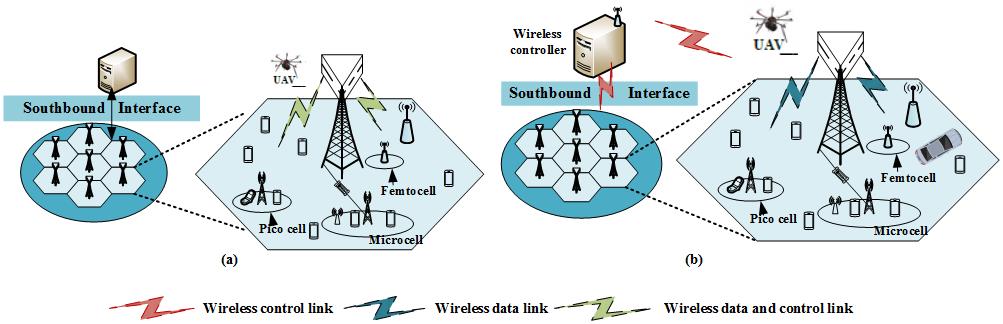}
\begin{figure*}[!htbp]
\centering \makeatletter\IfFileExists{81e5f09d-c93e-4c8f-8a10-0cdc71021460-ufig03_wcp_types.jpg}{\includegraphics{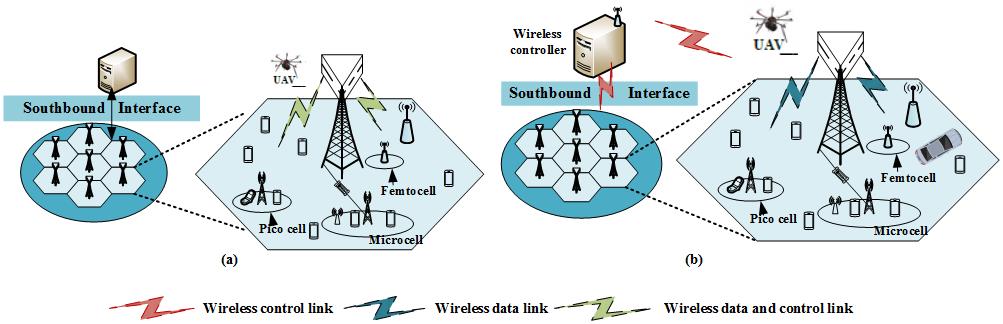}}{}
\makeatother 
\caption{{WCP-to-WDP communication scenarios: a) indirect WCP\unskip~\protect\citet{314005:7022495} ; b) direct WCP\unskip~\protect\citet{314005:7022256}}}
\label{figure-ff3b5984eac7508f369fbf1aa4dd8228}
\end{figure*}
\egroup
\mbox{}\protect\newline Without loss of generality, CRAN-based I-WCP and I-WCP fall in the same category (I-WCP) since CRAN-based indirect WCP has two CP domains (i.e., local control functions that are placed on the cloud of the CRAN wireless access network and the global control functions that are placed on the global SDN controller that has the global view of the whole network in wired and wireless domains). Similarly, CRAN-based D-WCP and D-WCP fall in the D-WCP category. \mbox{}\protect\newline Each category introduces an amount of CP delay that contributes to the ETE delay for 5G apps\unskip~\citet{314005:7022312}. In I-WCP, the CP's delay includes the SDN controller's queuing delay, the MWDP's queuing delay, a two-hop propagation delay, and a retransmission delay over the wireless channels. In D-WCP, the CP's delay comprises the SDN controller's queuing delay, the single-hop propagation delay, and the retransmission delay. In this sense, the I-WCP introduces two additional delay components, i.e., the MWDP's queuing delay and the MWDP-controller hop propagation delay. However, the latter could be ignored since fiber links would be deployed to connect an MWDP with an SDN controller.\mbox{}\protect\newline To explain the characteristics of each type, in the following subsection, a review of few use cases of WCP deployment in different 5G networks is discussed.

\subsection{The WCP use cases}Although the WCP design has not been discussed thoroughly in literature, it has been deployed in different 5G networks. Some of these networks have proposed the deployment of D-WCP, such as\unskip~\citet{314005:7022256,314005:7023627}. Another set has assumed the deployment of I-WCP, such as\unskip~\citet{314005:7022495}. Furthermore, there are some interesting 5G networks that have suggested the WCP deployment in their architecture but without identifying explicitly which category of the WCP they used, such as\unskip~\citet{314005:7022258}. Next, we discuss WCP deployment in a few 5G architectures: software-defined ultra-dense networks (SDUDNs), software-defined CRAN (SDCRAN), software-defined UAVs (SDUAVs), software-defined Internet-of-Vehicles (SDIoV), and software-defined IoT (SDIoT).

\subsubsection{SDUDN }In a UDN, the ultra-dense deployment of SBSs within the coverage area of the conventional MBS has been widely adopted for 5G networks to improve coverage and capacity. In\unskip~\citet{314005:7022256}, a cooperative multi-point (CoMP) communication scheme was proposed, where clusters of SBSs were created and the cluster's SBSs deploy CoMP communication to provide better coverage for many mobile users; thus, the MBS load was reduced (i.e., cell offloading). In this regard, an SDN solution was introduced to manage the SBSs' clustering process, where the SDN controller communicates directly with the underlying SBSs using wireless connections. In this system, a D-WCP was explicitly proposed, but the benefits of its deployment were not discussed. The potential benefits of D-WCP deployment in this network could be explained as follows.

\textbf{Lowering of the MBS's queuing delay --} To explain, assume that the I-WCP scheme is deployed. Therefore, the SDN controller and the MBS would process a huge number of control messages from a large number of SBSs, which is related to addressing legacy management (e.g., mobility management, spectrum sharing, and power management) in addition to the MBS's communication load and management of the SBSs' clustering process. However, in D-WCP, the MBS would not have the same control or communication load with the underlying SBSs as in I-WCP, and hence, the effect of MBS's queuing delay on the CP's delay is eliminated.

\textbf{Flexible assignment --} To illustrate, the unique feature of the SBSs, which is their operation in different power modes (i.e., on/off and active/sleep), is considered based on the communication activity of their underlying smart devices (e.g., mobile users' smartphones), to achieve the low power consumption requirement. Thus, the flexibility of using wireless links enables the dynamic assignment mechanism of the active SBSs to the wireless controller, and it contributes to achieving a scalable CP since a wireless SDN controller could be allocated to serve additional active SBSs if some of its assigned SBSs are in their sleep mode.

Nevertheless, the authors assumed that their D-WCP constitutes only a single wireless controller. According to their system model, the use of a single wireless controller could be acceptable, as they conducted their study on a single cell, which includes only one MBS along with the ultra-dense deployment of SBSs. However, it is not practically accurate to deploy a single wireless controller where the UDN has an enormous number of cells that are extended over a large geographical area. Thus, the deployment of multiple wireless controllers should be discussed to generalize this model. In this regard, the challenges of distributing these wireless controllers and wireless controller's carrier frequencies (i.e., allocating the spectrum band for control channels), deploying radio access technology, and addressing interference among the wireless controllers as well as with the WDP elements should be examined since they would degrade the potential improvements in scalability and flexibility performance when deploying D-WCP.

\subsubsection{SDCRAN }In\unskip~\citet{314005:7022265}, the radio and MAC functions were partially moved from the BBU to the RRH to accomplish a fine-grained decomposition for vBBSs. Instead of completely depending on the BBU to process all the radio processing and MAC functions (i.e., coarse-grained BS decomposition), RRHs can partially process these radios and MAC functions to offload the data exchange on the fronthaul network. In this network, SDN contributed to this fine-grained decomposition, where an CRAN-based I-WCP was proposed, by dynamically identifying a set of MAC and radio functions that should be moved from a vBBS at the BBU to a RRH. The controller utilized its global awareness of the wireless environment as well as the delay and throughput requirements for the 5G traffic to adaptively split the MAC and radio functions from the BBU and move them to the RRH to achieve fast signal processing. Although I-WCP was deployed in this system, D-WCP would be more appropriate than I-WCP. \mbox{}\protect\newline To clarify, a wireless controller that is placed in the vicinity of RRHs is assumed, and thus, it could periodically acquire the status of wireless channels. Furthermore, this wireless controller shares the global view of the whole network with a global centralized controller, which has the quality-of-service (QoS) requirements for 5G apps (e.g., the required ETE latency and throughput). Accordingly, the availability of the wireless channel status and QoS requirements would be utilized to adaptively optimize MAC and radio function splitting. By using I-WCP, RRHs need to impart the status of the wireless environment to a virtualized local SDN controller at the BBU to optimize the function splitting scheme. Clearly, the availability of the wireless status at a wireless controller would minimize the incurred delay of transmitting this information from the RRH to the SDN controller at the BBU, which also has to obtain the QoS requirements from the global SDN controller.

\subsubsection{SDUAV }A UAV (drone) is deemed to be the most crucial example that demonstrates why the WCP is needed, especially D-WCP. \unskip~\citet{314005:7022258} proposed UAVs that deploy different radio access technologies (RATs) to communicate with each other (e.g., LTE and WiFi), which creates a diverse range of wireless paths among the drones. This diversity in wireless paths introduces the challenge of how a UAV ensures reliable end-to-end connectivity with the other UAVs since they have limited onboard processing power that cannot afford frequent path computations due to their dynamic topology. Thus, the authors incorporated SDN into their system to provide a multi-path routing protocol to ensure resilient\footnote{Throughout the paper, the resilience of the system refers to the system or entity that intelligently resists any undesirable operational change.} end-to-end connectivity among the WDP elements (drones).

In this system, the SDN offered two major advantages: the WCP's global view of the drones' dynamic topology and the improved processing capability to perform path computation for the proposed multi-path routing algorithm. To capture the diversity of wireless paths, the proposed WCP obtained, through a dedicated control channel with each drone, the channel information from each drone in the DP on a real-time basis to build topology snapshots, and thus, the dynamics of the UAVs' topology state information were utilized by the SDN controller to reconfigure the multi-path routing protocol.

In this work, identifying the WCP category was overlooked. However, it can be understood from the context that it was a WCP since it is not practically feasible to connect drones with SDN controllers using wired links. Thus, there is a need to qualitatively evaluate the deployment of each WCP scheme in the context of this SDN-based system.

In I-WCP, an MWDP should be deployed to relay the control information to a centralized SDN controller. Intuitively, this MWDP should not be another drone since it would have the same limitations as the other drones. Fortunately, LTE was proposed (i.e., LTE drones) to provide a reliable command and control (C2) link for the safe operation of UAVs (drones)\unskip~\citet{314005:7022497}. Although LTE drones are in the nascent phase, considering an LTE BS as an MWDP for I-WCP deployment in SDUAVs could be realized.

In D-WCP, in a sense analogous to the first use case of the WCP in SDUDN, the D-WCP could be carefully designed to be incorporated into the UAV network. This potential deployment of D-WCP in SDUAVs would include the aforementioned potential benefit of lowering the MBS's queuing delay since the LTE MBS would not contribute to managing the drones. However, the challenge of distributing the wireless controllers in an efficient manner to fulfill the SDUDN and SDUAV design requirements still exists.

To mitigate the required transmission power of direct communication between an SDN controller and drones, \unskip~\citet{314005:7023627} discussed another SDUAV example in the context of tackling the controller placement problem (CPP). They assumed that the controller communicates with drones via a multi-hop communication method, where the controller is placed at the center of the UAV field and communicates directly with the closest UAV, which works as well as their neighbors as intermediate relay nodes between the controller and the remaining UAVs within the UAV DP. However, since the proposed topology assumed the deployment of a single controller, a discussion of the challenges of a potential link failure between the central UAV and any of remaining UAVs as well as of the UAV's capacity limitations is missing.

\subsubsection{SDIoV }SDN has been deployed in several vehicular network architectures, such as\unskip~\citet{314005:7022260,314005:7022320}. \mbox{}\protect\newline \unskip~\citet{314005:7022320} gave another example of an SDUDN, but in the context of a vehicular network. In this work, road offloading spots have been deployed as data exchange relays for connected vehicles. This model of connected vehicles has utilized the road infrastructure to offload the data exchange on the Internet infrastructure. An SDN controller has contributed to updating the offloading spots' flow tables and maintaining the status of the whole road networks. The proposed WCP type in this model was an I-WCP.

\unskip~\citet{314005:7022260} focused on optimizing the southbound communication between an SDN controller and the vehicular infrastructure (i.e., vehicular DP). A hybrid (vehicular ad hoc network, VANET, cellular-based) vehicular network scenario was proposed, in which I-WCP was deployed. The authors developed a game-based optimization model (i.e., rebating mechanism) for the WCP design to boost the latency performance for the critical safety applications in vehicular networks. In this model, vehicles transferred their control events to the controller through either cellular BS links or ad hoc links, whichever achieved minimum latency. The latency was interpreted in terms of deploying a rebating game to select the best link at a time slot that achieves low latency according to the rebated bandwidth game between the vehicles and the controller.

Although this system assumed I-WCP, incorporating D-WCP would be advantageous. With D-WCP, the links between the wireless controller and the underlying vehicles would participate in the proposed game. Additionally, D-WCP would not only contribute to managing the vehicle-to-vehicle (V2V) communication but also manage the vehicle-to-pedestrian (V2P) and vehicle-to-infrastructure (V2I) communication, which is called vehicle-to-everything (V2X)\unskip~\citet{314005:7022851}, where each part of everything is a smart wireless entity.

\subsubsection{SDIoT }Recent proposals for the Internet-of-things (IoT) deploy LTE systems to connect billions of smart things (i.e., smart sensors, smart homes, smart vehicles, etc.)\unskip~\citet{314005:7022254,314005:7022261}. In fact, the above-mentioned use cases are special examples of the IoT. \unskip~\citet{314005:7022261} introduced a thorough review of SDN deployment in the IoT and outlined its benefits, which is previously discussed; however, the discussion of the WCP design is missing in this work.

The deployment of D-WCP in IoT systems could be seen as more challenging than the indirect one since the IoT deploys different wireless technologies (e.g., ZigBee, WiFi and LTE). However, ongoing research efforts focus on converging the heterogeneous wireless technologies under the 5G umbrella, such as LTE deployment in unlicensed spectra\unskip~\citet{314005:7022849,314005:7022850}. Thus, LTE would be the 5G's backbone, and hence, the SDIoT architecture would be, in an analogous sense, the same as SDUDNs, SDUAVs, and SDIoV, but with a subtle change in the characteristics of the wireless devices and network applications. Therefore, D-WCP could be deployed in SDIoT networks as well as in the previous use cases.

The benefits and challenges of each WCP scheme are summarized inTable~\ref{table-wrap-f532a78fd7437dcf28bff66d131152a9}.

\begin{table*}[t!]
\caption{{ The WCP use cases.} }
\label{table-wrap-f532a78fd7437dcf28bff66d131152a9}
\def\arraystretch{1}
\ignorespaces 
\centering 
\begin{tabulary}{\linewidth}{p{\dimexpr.25\linewidth-2\tabcolsep}p{\dimexpr.25\linewidth-2\tabcolsep}p{\dimexpr.25\linewidth-2\tabcolsep}p{\dimexpr.25\linewidth-2\tabcolsep}}
\hline 
Architecture & WCP type & Benefit(s) & Challenge(s)\\
\tblmidrule 
SDIoV\unskip~\citet{314005:7022258}, SDCRAN\unskip~\citet{314005:7022265} &
  Indirect &
  No radical reform of the CP is needed &
  The DP is not completely decoupled from the CP since the WDP contributes to network orchestration.\mbox{}\protect\newline An undesired CP delay overhead is incurred due to the MBS's queuing delay and two-hop propagation delay between the global controller and SBSs, RRHs, RSUs, smart road infrastructure, or smart vehicles  \\
SDUDN\unskip~\citet{314005:7022255} &
  Direct &
  SDN vision was achieved by completely decoupling CP from DP.\mbox{}\protect\newline Lower MBS queuing delay due to relying on wireless controllers to process control events instead of the MBS.\mbox{}\protect\newline Flexible WCP-WDP assignment. &
  A single wireless controller was assumed, which is not adequate for large scale UDNs.\mbox{}\protect\newline Examination of the limitations of wireless channels, such as the interference effect on the WDP elements, is missing, which is a critical issue for addressing the scalability of the D-WCP\\
SDUAV\unskip~\citet{314005:7022256}, SDIoT\unskip~\citet{314005:7022260} &
  Not specified &
  By using I-WCP, only a soft implementation scheme would be needed since the WDP's resources would be utilized by the WCP to manage the 5G network.\mbox{}\protect\newline By using a D-WCP, a wireless controller would contribute to capturing the variations in the wireless channel and minimize the CP's delay overhead due to acquiring this information from the WDP elements. &
  For an I-WCP, a smart joint WDP and WCP resource management scheme is needed to fulfill the strict ETE QoS requirements for 5G apps (ETE delay requirement).\mbox{}\protect\newline For a D-WCP, effective distribution of wireless controllers along with achievement of the minimum delay and mitigation of the interference effect on both WDP elements and other wireless controllers are the major D-WCP design challenges.\\
\tblbottomrule 
\end{tabulary}\par 
\end{table*}
Clearly, both WCP categories face the limitation of exchanging critical control information within a wireless environment, which has a wide range of uncertainties (e.g., stochastic BS's load and stochastic retransmission) and its vulnerability to various types of attacks (e.g., software jamming and denial-of-service) in addition to ultra-dense deployment of heterogeneous entities.

\subsection{Related work}Although the WCP is in its infancy, there are a few proposals that discuss its design from the controller placement perspective, e.g.,\unskip~\citet{314005:7022498,314005:7022499,314005:7022406,314005:7022253}.

\unskip~\citet{314005:7022498} analyzed the uncertainty of the wireless environment by evaluating the quality of wireless links between the WCP (wireless controller) and the WDP (wireless BSs). In this work, D-WCP model was assumed where the quality of wireless links was estimated in terms of the number of retransmissions. Stochastic optimization tool\unskip~\citet{314005:7023676} was used to allocate the minimum number of controllers such that each base station (BS) \textit{b }gets a response from the associated wireless controller \textit{c} within a round-trip-time \textit{T\ensuremath{_{rtt}}} that is below a specific threshold,$\delta $. \textit{T\ensuremath{_{rtt}}} comprises wireless access delay \textit{T} (where time division multiple access, TDMA, is assumed), transmission and propagation delay$2{\tilde{n}_{bc}}t_{bc} $(where$ \tilde{n}_{bc} $is the number of transmissions\footnote{${\tilde{\cdot}} $ indicates that${\cdot} $is a stochastic variable.} that follows a geometric distribution), and controller's average queuing delay${\mathbb{E}}[T_{qc}] $. An equivalent deterministic mixed linear reformulation of the stochastic optimization model was solved to optimally distribute controllers such that the average number of transmissions satisfies the minimum delay requirement.

\textit{\unskip~\citet{314005:7022499}}discussed the uncertainty in the WCP design by studying the wireless controller placement where the WCP incurs chances of link failure, interference among wireless controllers, and additional interference between wireless controllers and wireless data plane since carrier sense multiple access with collision avoidance (CSMA/CA) protocol was assumed. In this work, the controllers were optimally distributed by identifying the minimum number of wireless controllers that are placed at certain locations and assigned to clusters of WDP elements where the quality of service (QoS) design requirements are satisfied. QoS metrics were defined as the probability of link failure, retransmission rate, average delay, and throughput. To solve this placement problem, a multi-objective optimization model was developed and solved by using Brute-force search algorithm.

\unskip~\citet{314005:7022406} studied the optimal distribution of the SDN controllers where each controller has an uncertain request rate due to the uncertainty of mobile user distribution within each cell. In this work, I-WCP scheme was assumed, and the mobile user distribution was studied to reflect the uncertainty of the WCP design. A two-stage stochastic optimization was used to solve the placement optimization problem where the objective is to minimize a linear combination of the number of controllers and the Round-trip-time \textit{T\ensuremath{_{rtt}}} (\textit{number of controllers}\textit{+}\textit{q}\textit{T\ensuremath{_{rtt}}}). To make the problem analytically tractable, authors only focused on the uncertainty in the controller's queuing delay, where each controller processes a different number of control requests from the assigned BSs; because the number of mobile users within each assigned BS is stochastic. They assumed that each mobile user has a constant traffic demand. The location of each user was modeled as a point that is placed according to the log-normal density function. The uncertainty in the mobile user distributions was modeled by developing 100 independent and identically distributed realizations (operational scenarios) of user distribution, each of 1000 users. The performance metrics were the number of controllers and the average probability of BS satisfaction\footnote{The probability of BS satisfaction is the relative frequency of realizations that satisfy delay requirement for the BS.} (averaged over all BSs).

Though the WCP is exclusive to the wireless networks, \textit{\unskip~\citet{314005:7022253}}discussed the WCP design for connecting data centers networks (DCNs). They discussed the design of WCP where mmWave wireless links between the SDN CP and DCNs data plane were used. In this work, the WCP design was also discussed by studying the placement of wireless controllers and developing a spanning tree routing algorithm for exchanging control messages over mmWave wireless links. The controllers were optimally placed where the minimum delay was achieved.

Although the aforementioned proposals for the WCP design attempted to address the challenge of uncertainty in the wireless environment, fitting all pieces of the WCP components in a robust framework is missing. Also, the discussion of WCP design from a single perspective (controller placement) is so inadequate that cannot effectively contribute to standardizing a design framework of WCP for 5G networks and beyond.\mbox{}\protect\newline 
    
\section{The Wireless Control Plane}
This section assesses the readiness of WCP deployment in 5G networks. First, the complexities of the WCP design, that are caused by the inherent limitations of the wireless environment (i.e., heterogeneity, uncertainty, and vulnerability), are discussed. Then, the WCP architectures are proposed along with examining their functionalities. Following that, a generic deep reinforcement learning (DRL) based WCP framework is presented, which would provide potential solutions for the WCP design challenges.

\subsection{The WCP's design complexities}The ultimate goal of the CP design is to obtain a reliable, scalable, flexible, resilient, secure, and green\footnote{Throughout the paper, the term green indicates that the system is sustainable, i.e., energy efficient.} WCP. Indeed, achieving these requirements for the WCP depends on the performance of the various entities within it. The reliability design issue, for example, can be discussed in the context of designing a reliable NOS, a reliable interface protocol, or a reliable controller's hardware. Similarly, the remaining design issues can be addressed from different perspectives. 

Most importantly, the trade-offs among these requirements bring the challenge of how to balance them to achieve the best performance of the WCP system\unskip~\citet{314005:7022263,314005:7023677,314005:7022500}. To illustrate, consider the reliability and green design issues that need to be addressed to ensure reliable delivery of the control information. The WCP might retransmit the control signals several times, which will consume additional power in communication and computing processes, and thus, attaining low power consumption might not be achieved.

Of course, most of the research efforts have been devoted to address the aforementioned design requirements in the case when the CP-DP connection is wired\unskip~\citet{314005:7022263}. Nevertheless, due to wide-ranging use of WCP for 5G networks and beyond, some of what have been discussed in Section 3, there is an urgent need to discuss these design issues from the WCP's perspective by considering the limitations of the wireless environment.

As discussed in Section 3, the WCP environment introduces three major challenges:

  \begin{enumerate}
  \item \relax Ultra-dense deployment of heterogeneous entities, where the WCP not only needs to manage heterogeneous BSs, RATs, and smart things but also must intelligently manage crowd sources of uncorrelated data\footnote{Sensed data from smart devices, smart vehicles, smart things, and even the feedback data responses from different software entities as well as the users' feedback.} to extract useful information that would contribute to making an optimum control decision.
  \item \relax The stochasticity of wireless cellular operation, which introduces different levels of uncertainty, such as:
  \end{enumerate}

\begin{itemize}
  \item \relax 

\begin{itemize}
  \item \relax The stochastic number of active wireless things\$ and, hence, the stochastic distribution of BS load.
  \item \relax The stochastic number of retransmissions.
  \item \relax The stochastic interference conditions.
  \item \relax The stochastic spectrum access schemes since 5G would rely on the coexistence of heterogeneous RATs. 
\end{itemize}
  
\end{itemize}
  This diverse range of uncertainties raises the challenge of adopting a WCP that has the capability to react effectively to random wireless operation circumstances.

  \begin{enumerate}
  \item \relax The vulnerability arising from the fact that SDN relies on software-defined modules. Therefore, the challenge of securing the WCP's software (NOS) against jamming and denial-of-service attacks\unskip~\citet{314005:7060336} as well as securing transmitted control signals over wireless channels should be addressed while designing the WCP.
  \end{enumerate}
  Accordingly, the WCP has to dynamically reacts to the stochasticity of wireless operation as well as maintains the integrity of the exchanged critical control information among crowds of heterogeneous entities to fairly satisfy the reliability, scalability, flexibility, resilience, security, and sustainability design requirements for the 5G critical applications.

Fortunately, deep reinforcement learning (DRL) has been recently utilized to tackle some of the sophisticated 5G WDP design challenges, e.g.,\unskip~\citet{314005:7022491,314005:7022492}. Thus, the DRL method could offer a holistic solution to tackling the WCP design intricacies. A DRL controller agent can successfully achieve a certain design goal or multiple design goals by learning from the controlled environment, which has crowd and heterogeneous sources of uncertainties\unskip~\citet{314005:7022311}. We refer to\unskip~\citet{314005:7022311,314005:7022491,314005:7022492,314005:7022493} for more details on the DRL topic and its recent advances.

Most importantly, at implementation time, there is a noticeable mutual benefit between the SDN and DRL. The SDN synthesizes a global view of the whole network by observing the status of all the network's entities, and thus, it benefits the DRL by providing it with a pool of data sources that interactively interpret the dynamics of the environment state. Indeed, this massive amount of collected data enables the DRL to make the optimum control decision precisely. Furthermore, it opens new horizons for online cross-layer optimization design in wireless networks. The DRL benefits the SDN through its flexibility to be implemented by different higher level languages (e.g., Java, C++ and Python), which facilitates its integration with the emerged SDN high-level programming language Pyretic\unskip~\citet{314005:7022317}.

The aforementioned discussion drives to two major methods for proposing a generic framework for the WCP design:

\begin{itemize}
  \item \relax DRL-based design, to benefit from the availability of a huge amount of system states at the central entity, which facilitates fast and accurate control decision-making. 
  \item \relax A programmable framework, to address the dynamic and stochastic variations in the wireless environment as well as provide a high-level programming environment for DRL implementation.
\end{itemize}
  Accordingly, in what follows, a programmable DRL-based framework for the WCP is proposed. The following subsection discusses the proposed DRL-based WCP architecture followed by a discussion of how does it work.

\subsection{WCP architectures}Since the proposed work concerns about examining the WCP design, in which the WDP-WCP connection is wireless, the proposed WCP architecture constitutes a model of the WDP as well as the WCP. Figure~\ref{figure-dd317350d859bcc575c5be2668f15305} depicts a generic model of the WCP, where the I-WCP and the D-WCP could communicate with the CRAN's WDP as well as the conventional cellular-based WDP architectures. Thus, the MWDP term is used to refer to the BBU in CRAN as well as to the MBS in the conventional cellular network. Similarly, the SBS term is used to refer to the WiFi AP, RSU, or UAV's communication payload.
\bgroup
\fixFloatSize{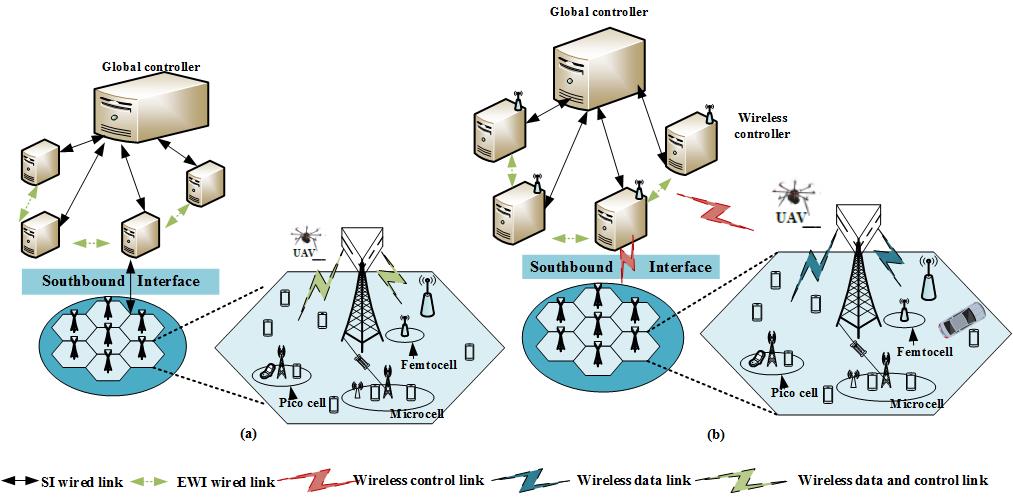}
\begin{figure*}[!htbp]
\centering \makeatletter\IfFileExists{91aed3e1-8999-4ff8-8074-f9794d30b359-ufig04_wcp_archits.jpg}{\includegraphics{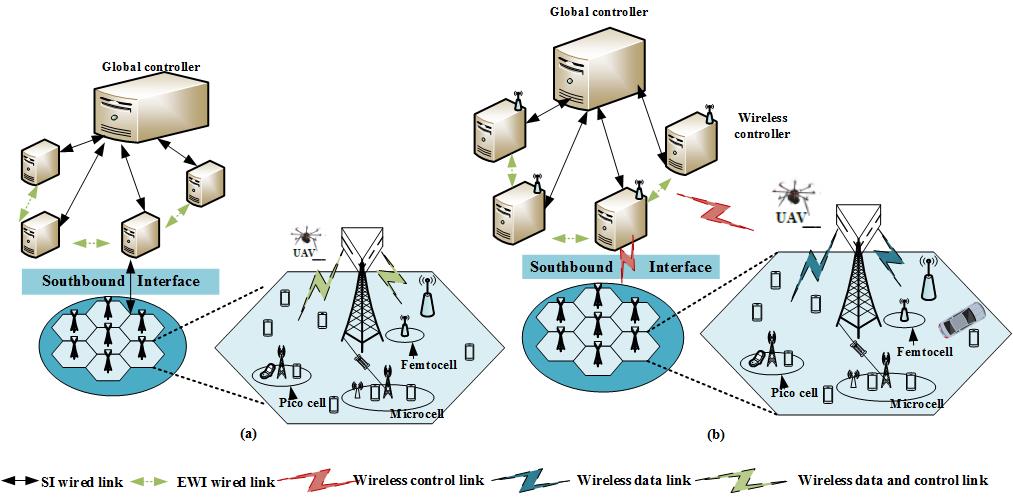}}{}
\makeatother 
\caption{{The WCP architectures a) I-WCP; b) D-WCP.}}
\label{figure-dd317350d859bcc575c5be2668f15305}
\end{figure*}
\egroup

\subsubsection{WDP's components}The deployment of a generic SDR-based programmable base station (PBS) model is assumed in both architectures except that the functionalities of the MWDP's control agent in the I-WCP's PBS are moved to the wireless controller in the D-WCP.

A PBS is proposed where its radio processing and waveform application modules provide a diverse range of RATs. Thus, the SDN controller has the capability to configure an off-shelf PBS via the PBS's communication service module (CSM) to enable any RAT (e.g., WiFi or LTE). The MWDP as well as the SBSs assume the deployment of the generic PBS model in their architectures with slight differences in the available radio resources and configuration methods for each type.

\bgroup
\fixFloatSize{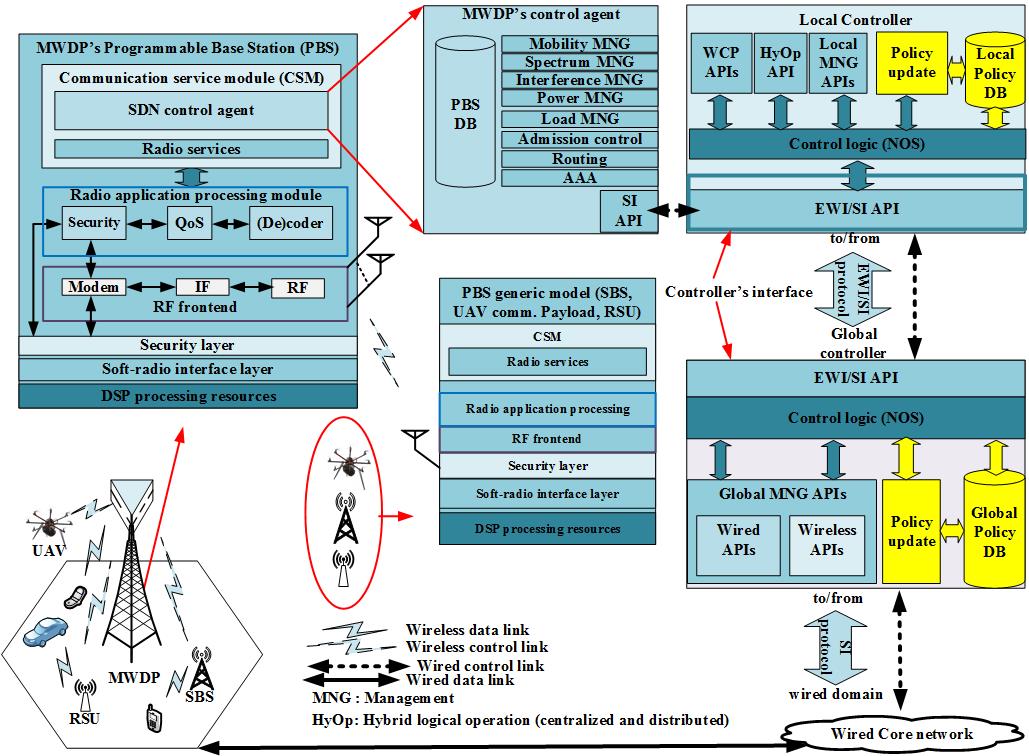}
\begin{figure*}[!htbp]
\centering \makeatletter\IfFileExists{13449ad2-36c4-491f-a0ae-30a06f337c9e-ufig05_iwcp_archit.jpg}{\includegraphics{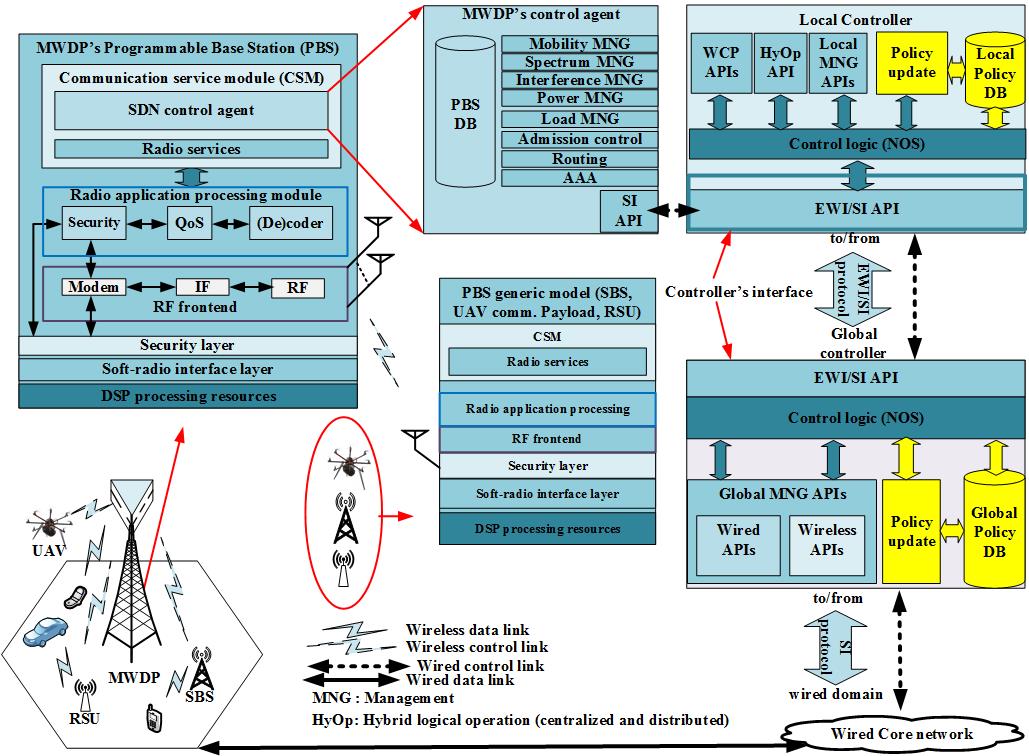}}{}
\makeatother 
\caption{{I-WCP functional block diagram.}}
\label{figure-4f16cedb859d863cd0f5f345e6216cc5}
\end{figure*}
\egroup

\bgroup
\fixFloatSize{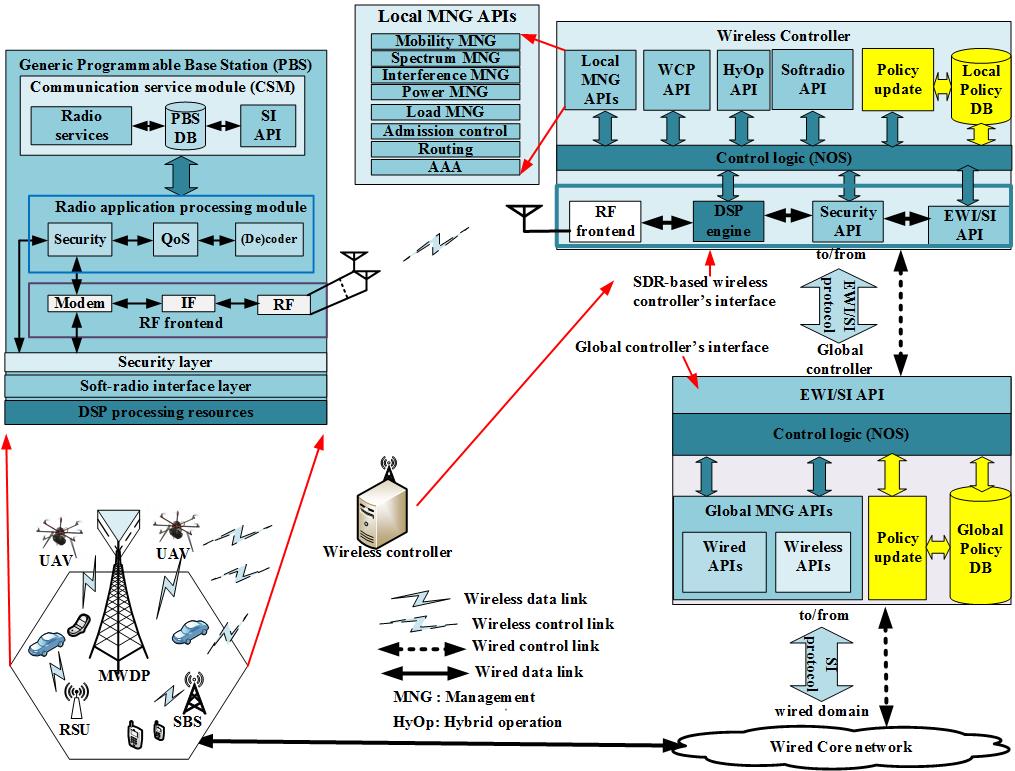}
\begin{figure*}[!htbp]
\centering \makeatletter\IfFileExists{f1bc5dc5-5c74-4c26-bb8d-0c4fc17f9477-ufig06_dwcp_archit.jpg}{\includegraphics{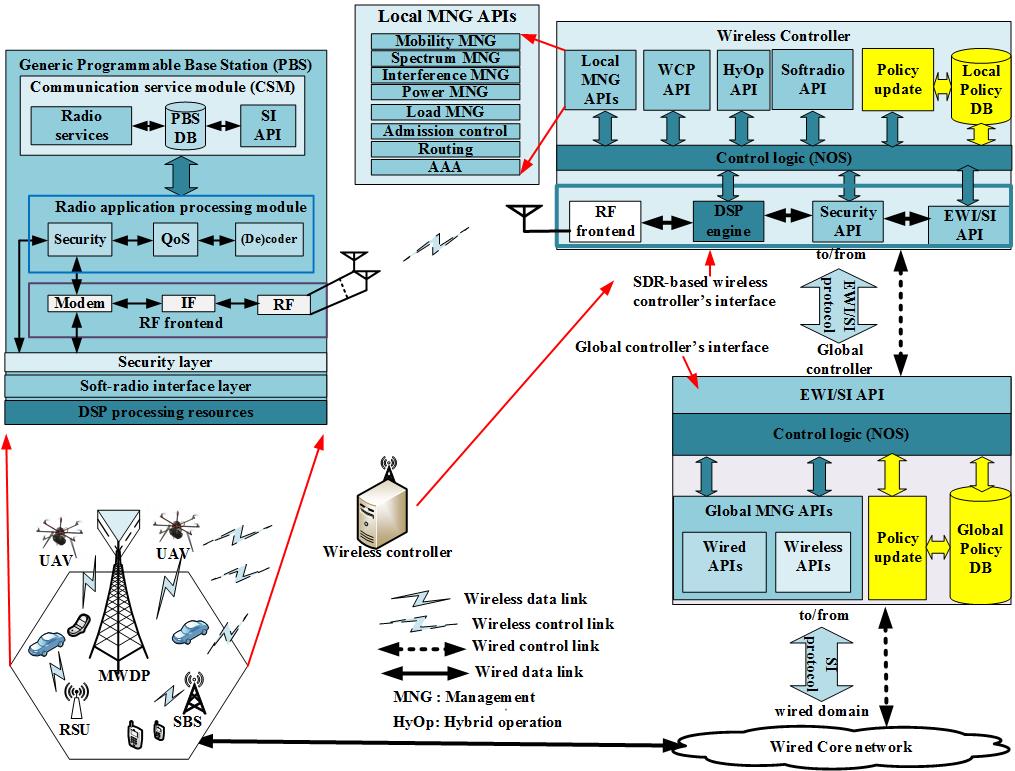}}{}
\makeatother 
\caption{{D-WCP functional block diagram.}}
\label{figure-d367bdee6cfddf21405f3e05ae6e0bb5}
\end{figure*}
\egroup
As shown in Figure~\ref{figure-4f16cedb859d863cd0f5f345e6216cc5}, the MWDP's CSM module includes the legacy SDR radio service module (e.g., spectrum sensing, authentication and necessary information for radio resource management) and the MWDP's control agent, while in the D-WCP, as shown in Figure~\ref{figure-d367bdee6cfddf21405f3e05ae6e0bb5}, the PBS's CSM includes only the legacy radio service module in addition to the SI agent. The SI agent is responsible for exchanging control information and updates between the CSM and the controller (e.g., spectrum sensing, hand-off and the PBS's load).

\subsubsection{WCP's components}As explained in Section 3, both WCP architectures assume the deployment of multiple distributed controllers\unskip~\citet{314005:7023675,314005:7023677,314005:7022498,314005:7022499,314005:7022500}. Furthermore, as shown in Figure~\ref{figure-dd317350d859bcc575c5be2668f15305}, both architectures deploy a hybrid topology that includes hierarchical along with cooperative topologies, in which distributed controllers are connected to a global controller, and they are capable of operating cooperatively to avoid a single point of failure (SPOF) problem and to attain a resilient and reliable WCP\unskip~\citet{314005:7022263,314005:7023677}.

Figure~\ref{figure-4f16cedb859d863cd0f5f345e6216cc5} depicts the I-WCP functional block diagram in which there are three control domains: the MWDP control agent, local domain, and global domain. The local and global control domains are genuine SDN domains, while the MWDP control agent is the SDN-like control unit within the WDP. The global domain manages the whole network in both the wireless and wired domains\unskip~\citet{314005:7023675}.

\textbf{MWDP's control agent:} This SDN-like unit has three basic modules:

  \begin{enumerate}
  \item \relax The database (DB) module, which maintain records of all information of the underlying wireless environment that the MWDP can provide to the local controller, such as:
  \end{enumerate}

\begin{itemize}
  \item \relax 

\begin{itemize}
  \item \relax Wireless channel states, by acquiring experienced throughput, number of retransmissions, or experienced delay.   
  \item \relax SBSs' load statistics. 
  \item \relax Request rate statistics. 
  \item \relax Information needed to build the interference map (i.e., by measuring the radio signal strength levels of the existing carrier frequencies).
  \item \relax Information that can be used by the MWDP itself to make a local decision, such as a spectrum sensing database, which would contribute to building an efficient dynamic access mechanism where different RATs coexist with each other (i.e., WiFi, LTE, radar)\unskip~\citet{314005:7022495}).
\end{itemize}
  
\end{itemize}

  \begin{enumerate}
  \item \relax The control processing module, which utilizes the local information in the DB to make a local control decision (e.g., spectrum switching\unskip~\citet{314005:7022495}). 
  \item \relax The SI interface module, which deploys an SI protocol (e.g., OpenFlow) to encapsulate/decapsulate the control information that is exchanged with the local controller\unskip~\citet{314005:7040583}.
  \end{enumerate}
  \textbf{Local controller:} There are three major entities of the local controller:

  \begin{enumerate}
  \item \relax The controller's computing, memory, and storage resources.
  \item \relax Control logic (NOS), which is the controller's backbone and not only manages the controller's resources (i.e., computing, memory, APIs, etc.) to provide the MP with the control development environment\unskip~\citet{314005:7022318} but also, as it will be discussed later, represents the DRL environment of the proposed DRL framework for the WCP design.
  \item \relax  Control APIs:
  \end{enumerate}

\begin{itemize}
  \item \relax 

\begin{itemize}
  \item \relax Replicated MWDP control units but at a broader scale to manage several cells (MWDPs with their underlying SBSs), which are assigned to them.
  \item \relax An application module responsible for (re)configuring policies in the wireless domain (e.g., routing, radio resource management, mobility management, power and spectrum)\unskip~\citet{314005:7022495,314005:7023675,314005:7022265}. 
  \item \relax The WCP API, which its design framework will be discussed in the following subsection.
\end{itemize}
  
\end{itemize}
  \textbf{Global controller: }It is located at the top of the WCP hierarchy. It includes the global modules of the local controllers in addition to its global control APIs to manage both the wired DP and WDP devices. Additionally, it includes a global policy database along with a global policy updater that provides the necessary cross-layer information for ETE performance optimization purposes\unskip~\citet{314005:7022495,314005:7022265,314005:7023675}. 

In a D-WCP, as shown in Figure~\ref{figure-d367bdee6cfddf21405f3e05ae6e0bb5}, a framework for a wireless controller design is presented. The MWDP's control agent APIs are moved to the wireless controller. Furthermore, an SDR-based programmable wireless interface is embedded in the proposed wireless controller model.

\textbf{The wireless controller:} It resembles the local controller structure in I-WCP, but in addition to the transferred MWDP's control agent functionalities, it has two distinct modules:
  
  \begin{enumerate}
  \item \relax An SDR-based wireless interface that includes the legacy SDR components in addition to two important software modules:  
  \end{enumerate}

\begin{itemize}
  \item \relax 

\begin{itemize}
  \item \relax A security agent that offers the opportunity to achieve robust security for the WCP.
  \item \relax An east-west interface and southbound interface (EWI/SI) agent that manages not only the communication between the controller and the PBS but also the communication between it and its counterparts as well as the global controller.
\end{itemize}
  
\end{itemize}

  \begin{enumerate}
  \item \relax A soft-radio control API, which provides the development environment for the SDN network developer to not only configure the SDR operation but also directly program the SDR entities in the underlying WDP (i.e., integrating the two SDN and SDR programming environments). Section 5 discusses this issue in more detail.
  \end{enumerate}

\subsection{DRL-based WCP framework}In a DRL, the agent (control API) applies policy ($\pi $) to an observed state (\textit{s\ensuremath{_{t}}}) and returns an action (\textit{a\ensuremath{_{t}}}) to the environment (control logic), which changes its state to another one and returns its new state (\textit{s\ensuremath{_{t+1}}}) along with an immediate reward (\textit{r\ensuremath{_{t}}}) to the DRL agent. (\textit{s\ensuremath{_{t+1}} , r\ensuremath{_{\textit{t}}}}) are used for policy optimization purposes by using different learning methods, such as the deep Q network (DQN), which is based on the convolution neural network (CNN)\unskip~\citet{314005:7022493}. The DQN estimates the expected long-term reward ($Q_{\pi}(s_{t}, a_{t}) $) of the consequent state-action pairs ($\{ (s_{t+1}, a_{t+1}), (s_{t+2}, a_{t+2}), ... , (s_{t+n}, a_{t+n}) \} $) if the policy ($\pi $) applied. This long-term reward interprets the achievement of the design goal. 

Of course, there are different design issues that rely on several WCP entities; however, the DRL has two major features (i.e., recursive and cooperative). DRL can be realized recursively\unskip~\citet{314005:7022311}, which the DRL framework could be represented in its base case, and then, a robust DRL framework could be synthesized recursively. Additionally, DRL can cooperate with other DRL-based control systems to achieve complex design goals. In this regard, a centralized DRL-based SDN control plane would include different DRL control APIs that share a global control environment (i.e., controller's operating system). Thus, each DRL-based control API (DRL agent) would share the control environment (controller's operating system) with the other DRL APIs (other DRL agents) where each has policy $\pi $ that could be adapted according to instantaneous observations $o_{t} $ from the surrounding control environment. Thus, a specific DRL-based control API (e.g., controller placement) could observe the instantaneous state $s_t $ of another DRL control API (e.g., security API) to achieve certain design goals (e.g., scalability and security). Accordingly, the DRL agent observes the instantaneous changes in the surrounding control environments, which could indicate to a change in the reliability, scalability, flexibility, resilience, security, and/or sustainability design metrics and assess (through reward signals) whether its policy $\pi $ fulfills these design requirements or not.

To exemplify, the wireless SDN controller constitutes the following DRL-based control APIs: controller placement API, security API, power management API, interference management API, admission control API, and software management API. Indeed, each agent of these DRL control API shares the observations of the other APIs since they have the same control environment (i.e., controller's operation system). Nevertheless, each agent should utilize a specific set of available observations that would contribute to achieving a specific design goal. To explain, assume the design of DRL-based controller placement API, which could cooperate with the other APIs to achieve a certain design goal. Assume the goal is to fairly satisfy the reliability, scalability, flexibility, resilience, security, and sustainability design requirements for the 5G critical applications.\mbox{}\protect\newline Undoubtedly, balancing the aforementioned design requirements is not a trivial task. However, the potential gain of using cooperative DRL APIs within a global control environment would balance these design requirements.

In this regard, the controller placement API would cooperate with all of the other APIs to optimally assign and re-assign controllers to WDP elements such that the required design goal could be achieved. Thus, DRL's agent of the controller placement API would observe the designated observations of each DRL's agent of the following control API:

\begin{itemize}
  \item \relax Security API, which could contribute to assigning the controllers to WDP elements where the control messages are securely exchanged (security requirement), 
  \item \relax Power management API, which could guarantee that the controllers are placed and assigned to WDP elements where the power consumption at the lowest possible level to attain the sustainability design requirement, 
  \item \relax Interference management, which could contribute to achieving:

\begin{itemize}
  \item \relax The flexibility design requirement, since the interference levels are changing due to the stochasticity of wireless environment, flexible and dynamic assignment/reassignment between the wireless controller and the WDP elements could be achieved where interference could be avoided (i.e., this introduces positive effects on the number of retransmissions, and thus, low power consumption and low latency design requirements could be satisfied),
  \item \relax The scalability design requirement, as interference affects the number of retransmissions, the capacity of controllers' buffers will be rapidly occupied as long as the number of retransmissions grows, and hence, the incoming control requests from other WDP elements will be dropped out. Thus, when interference avoidance is achieved, the scalability of WCP will be elevated,
  \item \relax The reliability design requirement, similarly, as interference increases, the reliability of exchanging control messages will be degraded
\end{itemize}
  
  \item \relax Admission control API, this API's observations could contribute to positioning and assigning the controllers to the WDP elements where controllers' capacities could afford (admit) as many incoming requests as possible from the associated WDP elements (scalability design requirement), 
  \item \relax Software management API, its observations could contribute to placing the controllers and assigning them to the WDP elements such that SDN's software system could be self-healed (resilience requirement)
\end{itemize}
  Accordingly, the controller placement DRL's agent observes the instantaneous changes in the surrounding control environments (i.e., observe the security API's changes due to a potential malicious activity, or observe the interference API's changes due to the stochasticity of the interference map). The aforementioned instantaneous observations from all surrounding APIs would indicate a change in the reliability, scalability, flexibility, resilience, security, and/or sustainability design goals.

Thus, by assessing the controller placement API's policy $\pi $ (through reward signals), it can be adapted according to an instantaneous observation $o_t $ from the surrounding control environment as well as from the other APIs, which each observation interprets the DRL's environment instantaneous state $ s_t $.

The aforementioned example emphasizes the significance of designing the observation signals $O_{t} $ for the DRL environment.

Most importantly, the DRL's environment is deemed to be its all surrounding entities whose states need to be observed. These entities might also be DRL-based entities that observe this entity's states (e.g., DRL controller placement API's agent would observe the DRL-based interference, power, and load distribution environment states, and similarly, each of their agents would observe the DRLWCPP's environment states). Thus, it is critically important to define the domain of observations that the control agent needs to observe to take appropriate action to achieve the design goals. A discussion of this issue as well as other important design aspects is included while describing the proposed DRL-based WCP.

\bgroup
\fixFloatSize{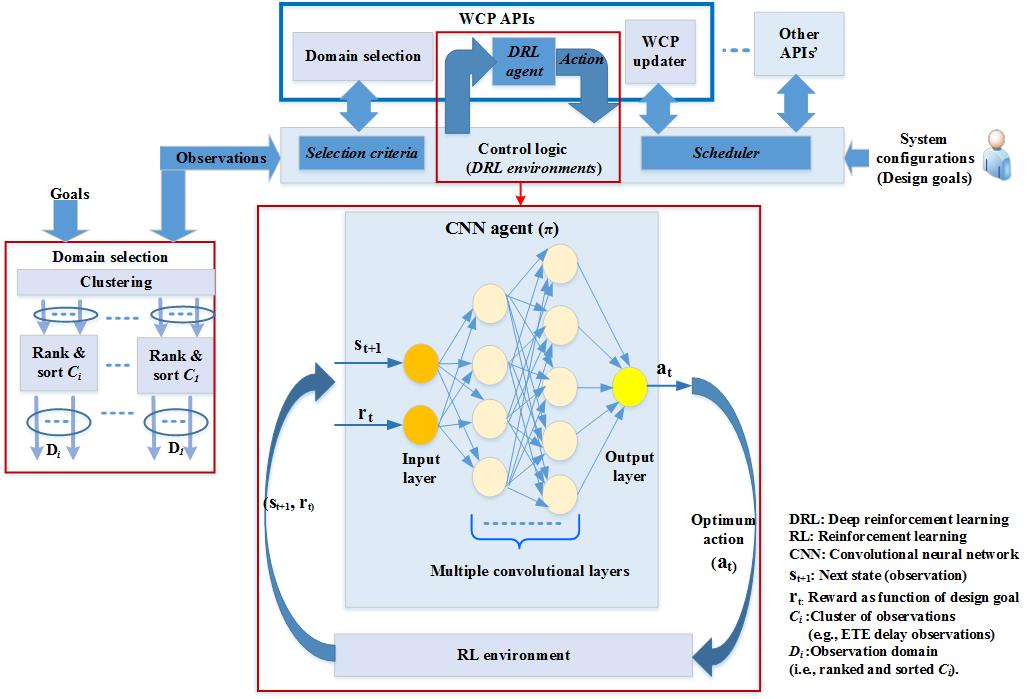}
\begin{figure*}[!htbp]
\centering \makeatletter\IfFileExists{5e622271-aa78-4f9a-808c-4dc4647f4940-ufig07_wcp_framework.jpg}{\includegraphics{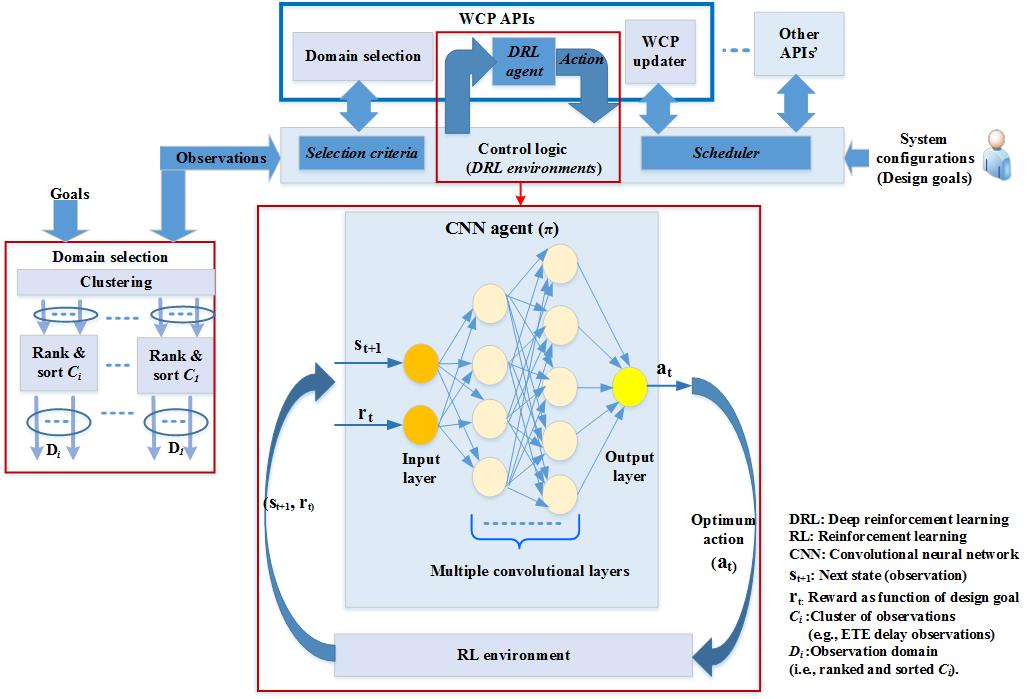}}{}
\makeatother 
\caption{{The WCP functional block diagram.}}
\label{figure-4b3af5538334c2e17b549906aec8681a}
\end{figure*}
\egroup

\bgroup
\fixFloatSize{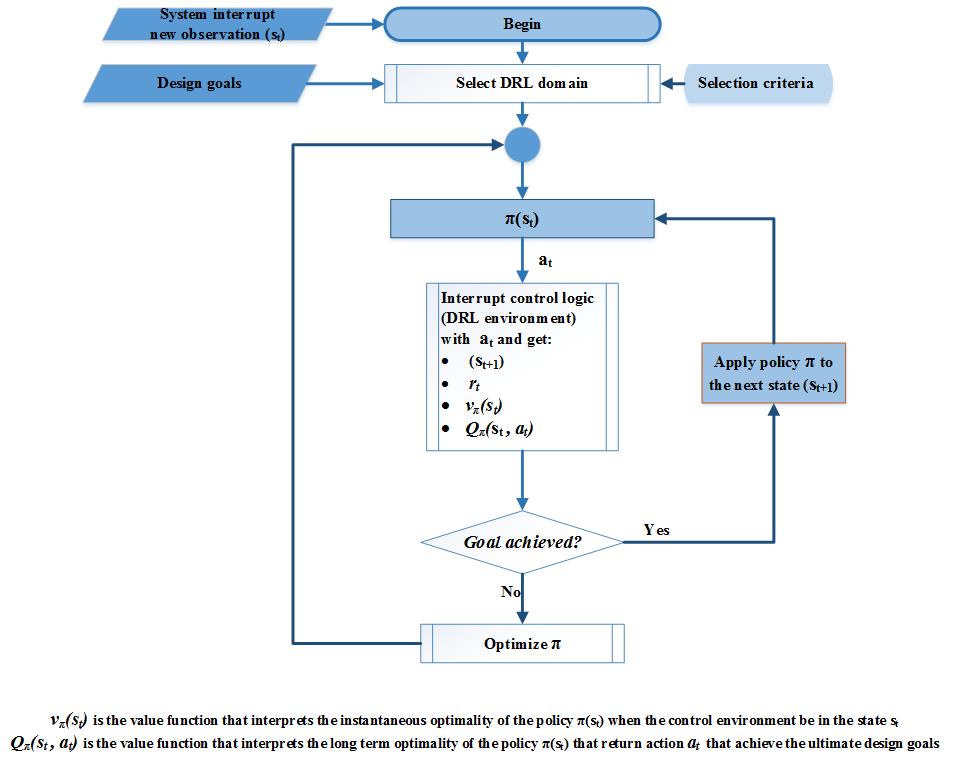}
\begin{figure*}[!htbp]
\centering \makeatletter\IfFileExists{d99ea417-a648-43da-b12a-9d56f6588065-ufig08_wcp_workflow.jpg}{\includegraphics{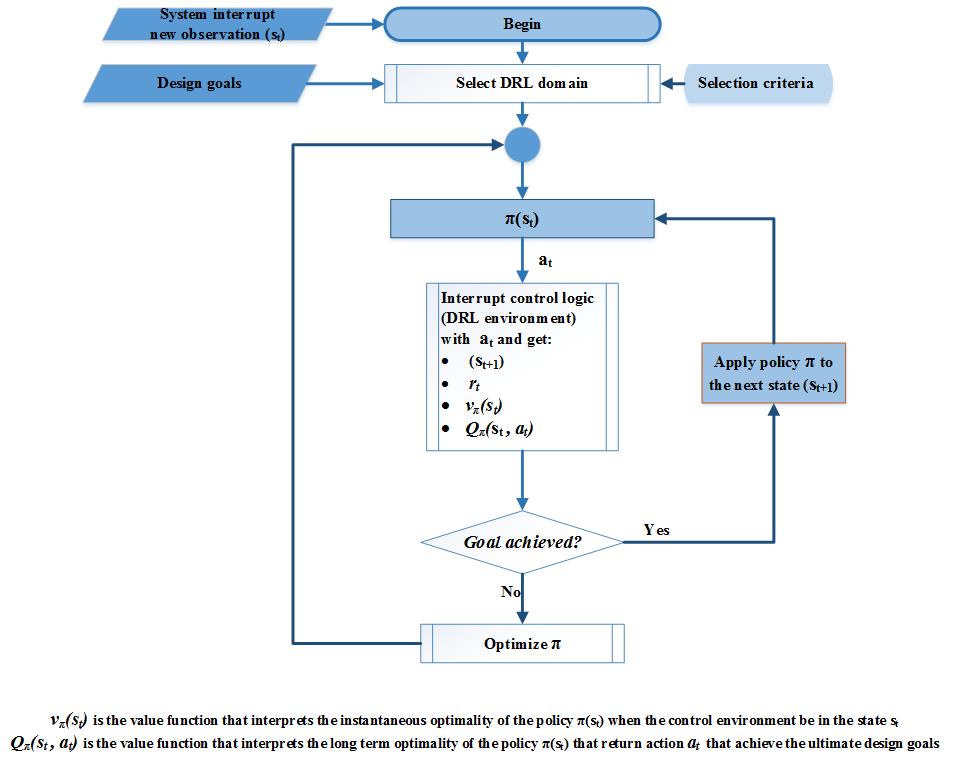}}{}
\makeatother 
\caption{{The WCP workflow.}}
\label{figure-3c9a482b802102e0a7f1919c1974b217}
\end{figure*}
\egroup
A generic DRL-based WCP framework is depicted in Figure~\ref{figure-4b3af5538334c2e17b549906aec8681a}, and its workflow is demonstrated in Figure~\ref{figure-3c9a482b802102e0a7f1919c1974b217}. The WCP API's DRL agent (i.e., a convolution deep neural network) interacts with the DRL environment, which is the controller's OS or control logic that manages all the controller's resources. The DRL environment observes all states from all control entities. It is assumed that other APIs replicate the same structure but not the same functions of the WCP API, in which all are implemented by DRL methods.

The DRL-based WCP process has three stages: preprocessing, processing, and post-processing.

\subsubsection{Preprocessing}In preprocessing (i.e., the offline learning stage) SDN and DRL developers would perform the following processes for each DRL-based API.

\textbf{Identify design goals:} Recall the DRLWCPP example; observations of security system behavior become required when the security design goals are identified. Thus, identifying the design goals is the key to DRL. In the proposed framework, design goals can be easily configured by the network developer or operator, who can utilize the SDN and SDR programmability to attain this objective.

\textbf{Identify the set of observations:} The set of observations $O_t $, which is represented by the domain selection module as shown in Figure~\ref{figure-4b3af5538334c2e17b549906aec8681a}, is identified, in which the following subprocesses are executed:

  \begin{enumerate}
  \item \relax The SDN controller's logic obtains updates from all network entities, and these updates represent DRL observations. 
  \item \relax Since the controller receives observations from crowd entities, a clustering algorithm is applied to identify the most important observation that each DRL agent should handle according to a specific design metric \textit{i} where $i \in I $ and$ I = \left\{ETE_{latency}, ETE_{throughput}, \cdots \right\} $.  $I $is a set of design metrics that could contribute to clustering the observations according to a designated design goal (e.g., the set of observations that reflects the ETE latency contributes to achieving the ultra-low latency design goal for 5G critical applications).
  \item \relax In each cluster $C_{i}, C_{i} = \left\{O_{t_{i}} | i \in I \right\} $ the observations are ranked and sorted according to a specific criterion (i.e., the ranked and sorted set of each observation cluster $ C_{i} $ is defined by the DRL domain $D_{i} = \left\{ranked\space and\space sorted\space C_{i}\space \forall\space i \in I \right\} $, such as ranking the observed states from the worst to the best according to observed ETE performance (i.e., the selection criteria can be configured in the selection criteria module in the controller's logic). This process would solve the problem of learning the DRL's DQN with data that are parametrically correlated but in reality might have a different interpretation of observed information (i.e., back-propagation with a stochastic function was recently proposed to overcome this limitation of the DQN-based reward estimation\unskip~\citet{314005:7022493}.
  \end{enumerate}
  \textbf{Identify the set of all possible actions: }This set represents the DRL agent's output.

\textbf{Design state signal:} It interprets an observation $o_{t} \in O_{t} $ at each time step \textit{t} to a state signal \textit{s\ensuremath{_{t}}} (control system interrupt) that invokes the DRL agent process.

\textbf{Design immediate reward signal: }This signal, \textit{r\ensuremath{_{t}}}, is a function of one or more of the design goals\unskip~\citet{314005:7022311,314005:7022493}.

\textbf{Design value function:} There are two types of value functions:

  \begin{enumerate}
  \item \relax 
  
  \begin{enumerate}
  \item \relax Instantaneous value function $v_{\pi}(s_{t}) $, which interprets the instantaneous optimality of the policy $\pi(s_{t}) $ when the control environment be in the state \textit{s\ensuremath{_{t\ }}}
  \item \relax Long term value function $Q_{\pi}(s_{t}, a_{t}) $, which reflects the optimality of applied policy $ \pi $, in the long term, that returns actions where ultimate design goals can be achieved.
  \end{enumerate}
  
  \end{enumerate}
  As mentioned earlier, different DQN schemes with different types of approximation techniques have been proposed to address the $Q_{\pi}(s_{t}, a_{t}) $ estimation, which is still an open design issue for further research.

\textbf{Environment model:} DRL was originally devised to react to the dynamics of the environment in real-time based on a trial-and-error learning strategy. However, the challenge of designing a stochastic DQN estimation function as well as the complexities of the agent's surrounding environments could result in high computational complexity when executing the DRL algorithms in real-time using the trial-and-error learning scheme. In this regard, data scientists have recently devoted efforts to examining different stochastic tools (e.g., stochastic geometry\unskip~\citet{314005:7022494} and stochastic optimization\unskip~\citet{314005:7023676}) to model the surrounding environment for DRL operation\unskip~\citet{314005:7022493,314005:7022311}. Accordingly, this would provide the capability to expect the environment's behavior when applying different actions to it, and hence, this approach would abate the DRL's computational complexity during real-time operation by using a predefined stochastic model that could reflect the system uncertainties\unskip~\citet{314005:7022311,314005:7022493}.

\subsubsection{Processing}As shown in Figure~\ref{figure-4b3af5538334c2e17b549906aec8681a} and Figure~\ref{figure-3c9a482b802102e0a7f1919c1974b217}, control logic would utilize the configured design goals and selection criteria that are embedded within the control logic to set the domain of observed data for each DRL-based control API. Recall the benefit of applying SDN-centralized control logic to DLR that the trajectories of previous observations from all control APIs and ETE performance updates could be used to set up a precise observation domain for each DRL-based control API. In fact, the domain selection module could be constructed by DRL methods since it has shown noticeable success in robotics by enabling them to recognize many complex patterns\unskip~\citet{314005:7022493}. Thus, clustering, ranking, and sorting algorithms could also be represented by DRL methods\unskip~\citet{314005:7022311}.

After selecting the domain for each DLR-based control API, Figure~\ref{figure-3c9a482b802102e0a7f1919c1974b217} abstractly demonstrates the workflow of a single DRL API. It initializes the operation by applying the policy $\pi $ to the first stimuli \textit{s\ensuremath{_{0\ }}}; $a = \pi(s) $. Then, it responds by deciding to apply an action \textit{a\ensuremath{_{0}}} to the control logic, which applies it to the appropriate control APIs. Following that, the control logic observes (senses) the next state \textit{s\ensuremath{_{1}}} and returns it with an immediate reward \textit{r\ensuremath{_{1}}} to the agent. Concurrently, the control logic estimates the DQN long run reward \textit{Q(s, a)} that is formally defined in\unskip~\citet{314005:7022311}, $Q(s, a) = ((1-\alpha) \times Q(s_{t}, a_{t})) + (\alpha \times (r_{t} + \gamma \times \max\limits_{a}Q(s_{t+1}, a))) $, where $\alpha $ is the learning rate, $ \gamma $ is the discount factor, and $\max\limits_{a}Q(s_{t+1}, a) $ is the estimate of optimal future reward values. 

Estimating the optimum value of expected rewards is NP-hard; therefore, a wide range of approximation algorithms have been proposed to address this problem\unskip~\citet{314005:7022311,314005:7022493}. The proposed framework could deploy the duel deep Q neural network learning method (DDQN)\unskip~\citet{314005:7022496} for Q estimation. \mbox{}\protect\newline The potential deployment of the DDQN is motivated by the following facts:

\begin{itemize}
  \item \relax Its successful deployment in a real world problem (i.e., pattern recognition for robotics)\unskip~\citet{314005:7022311,314005:7022493}. 
  \item \relax Its approximation hypothesis that lies in the fact that estimating the value of each action \textit{a }for a given state \textit{s} is not crucial because there are several states for which the policy does not need to choose an action to respond to it, in other words, many states need only a single action from the agent (policy) to respond with\unskip~\citet{314005:7022496}.
\end{itemize}
  To exemplify the second motivation of DDQN deployment in our WCP world, consider that the proposed DRLWCPP observed a malicious state. In this sense, there is not a pool of choices for the agent to respond with one of them. It is only the action of disinfecting the system software that the agent must respond with. Thus, estimating the expected rewards for each state-action pair should be weighted with an advantageous factor that would contribute to accurately approximating the solution of the \$Q\$ maximization problem. \mbox{}\protect\newline Accordingly, the authors in\unskip~\citet{314005:7022496} decomposed $Q(s, a) $ into two separate functions: $Q_{\pi}(s_{t}, a_{t}) =  V_{\pi}(s) + A_{\pi}(s, a) $, where $V_{\pi}(s) $ is the value function and $A_{\pi}(s, a) $ is the advantage function. In this regard, two streams of fully connected layers, which share the same convolutional learning module, have been deployed to provide separate estimates of the value and advantage functions. Then, the outputs of these streams are merged to provide a single output of the \textit{Q} estimate. A detailed description of the estimation procedures and experiments are described in \unskip~\citet{314005:7022496}. 

The following process in DRLWCPP is used to verify if the design goal has been achieved for the given policy $\pi $. If yes, continue the DRL operation with the consequent state-action pairs. If no, optimize the CNN policy. The policy could be optimized by using the stochastic value gradient (SVG) method that was demonstrated by \unskip~\citet{314005:7022309}. The incentives to incorporate this optimization method into the proposed framework are as follows:

\begin{itemize}
  \item \relax It has achieved noticeable improvement in complex control systems (e.g., real visual navigation tasks) \unskip~\citet{314005:7022310}. 
  \item \relax It combines the powerful features of the value gradient method (i.e., optimizing CNN by the backpropagation learning method) and the stochastic learning method (i.e., optimization relies on the likelihood ratio estimator of the return reward samples from the environment). 
\end{itemize}
  To illustrate the significance of last incentive, consider the proposed DRLWCPP framework in which critical control actions have to be executed in efficient timely and reliable manners. Assume that, in real-time operation, policy $\pi $ of a DRL agent returned an action to the control logic that did not achieve a design requirement; thus, a fast and reliable online optimization scheme is required. \mbox{}\protect\newline Accordingly, using only the value gradient backpropagation optimization method (i.e., minimizing the error between the required CNN's output and the obtained feedback by adjusting the CNN's weighting factors) is not effective since it works for optimizing deterministic policies. Moreover, the deployment of only the stochastic learning method, which relies on estimating the probabilistic ratios of samples (samples of DRL's environment returns), lacks accuracy due to suffering from high variance caused by estimating the average values of divergent samples. In this sense, the availability of a huge number of samples would minimize the variance, and hence, improve the efficiency of this technique. Thus,  \unskip~\citet{314005:7022309} developed a stochastic value gradient method to optimize the stochastic policy by using real trajectories (a huge number of samples) that are stored in a system database to train the CNN agent (i.e., this recalls the major benefit of SDN for online optimization through the use of datasets from the real environment instead of relying on expected data to train the stochastic policy). Accordingly, the CNN agent's policy ($\pi $) is stochastically trained on real-world datasets that interpret the uncertainties of the controller environment. Hence, the policy could respond optimally, in real time, to stochastic events and could effectively achieve the targeted performance. We refer to \unskip~\citet{314005:7022309}, for more details about the implementation of the stochastic value gradient (SVG) optimization method.

\subsubsection{Post-processing}In this stage, all trajectories of state-action pairs as well as overall observed instances of system performance could be stored and backed up for further data analysis. As explained above, the power of the estimation methods lie in using datasets from the real world. Additionally, the availability of this type of data would help researchers, in academia and industry, from different areas (data science, stochastic optimization, stochastic geometry, digital signal processing, wireless communication, information theory, system engineering, network programming, network management, software engineering, intelligent transportation systems, risk management, etc.) to learn about the effectiveness of their methods and how they are so synergistic that they can serve the whole of mankind.
    
\section{Is the WCP Ready for 5G Networks and Beyond? Directions for Future Research}
The previous sections have discussed what the wireless control plane is, why it is significance for 5G networks and beyond, how it is proposed for managing different 5G architectures, and what design challenges it faces. Then, a framework that would pave the way for implementing an efficient WCP has been proposed. Nevertheless, the proposed WCP is in its infancy, and there are many interesting and challenging research problems that need further efforts. Therefore, in what follows, a few of these challenges that are open for further investigations are highlighted.

\subsection{Direct or indirect WCP}\textbf{Challenge:} As demonstrated in Section 3, each WCP scheme has its benefits and limitations that necessitates further investigation to identify when and how the deployment of one of them would be more beneficial than the other one. It has been stated that the D-WCP would offer lower latency than the I-WCP because the WDMP's queuing delay would be eliminated. Although this hypothesis might be qualitatively true, an experimental study is needed to compare the performance between each scheme in different scenarios. In particular, addressing the flexibility and reliability trade-offs in D-WCP by studying the effects of the wireless controller's interference on the number of retransmissions when a controller decides to assign another WDP at the edge of its coverage range. This type of retransmission's delay overhead, due to enabling flexible assignment, needs to be quantitatively compared with the MWDP's queuing delay in the I-WCP scheme. To conduct this comparison thoroughly, the problem of wireless controller placement should be revisited to address the effect of interference, from WDP elements as well as the other wireless controllers, on the reliability of the WCP. Although a few research efforts have addressed the wireless controller placement problem (WCPP)\unskip~\citet{314005:7022498,314005:7022499}, a comparison with I-WCP is missing.

\textbf{Potential approach:} To perform a comprehensive comparative study between I-WCP and D-WCP, incorporating the WCPP into the WDP placement problem could be examined. The WDP placement problem, such as the placement problem of mmWave SBSs in UDNs to improve coverage performance, is exhaustively researched by using stochastic geometry, stochastic optimization, and game theory analytical tools\unskip~\citet{314005:7022315,314005:7023585,314005:7022316}. In this regard, a joint placement optimization framework could be devised to address the challenges of the WCPP and the WDP placement. Thus, the problem of how the controllers and SBSs would be jointly distributed and dynamically assigned to each other in an optimal way to achieve better coverage and provide URLLC services for 5G systems should be addressed. On the other hand, the CPP problem for I-WCP should also be studied thoroughly to address the uncertainties of the wireless environment in the WDP. \unskip~\citet{314005:7022406} took a step in this direction by addressing the uncertainty of the BS load distribution due to the stochasticity of the mobile users' activity within each cell. However, there are still different aspects that need to be examined, such as examining the placement problem where two sources of uncertainties are present (e.g., the uncertainty of the BS load distribution in addition to the uncertainty of retransmission rate).

\subsection{Securing the WCP control}\textbf{Challenge:} The WCP system has three major critical design issues:

  \begin{enumerate}
  \item \relax Sensitive control information is prone to attack due to the distributed processing for control events among the distributed controllers. 
  \item \relax The entire operation relies on susceptible software-based models.
  \item \relax Sensitive control information is broadcast over a vulnerable wireless channel.
  \end{enumerate}
  All of these design aspects raise the chances of attacking the WCP, e.g., denial of service (DoS), distributed denial of service (DDoS), and software jamming\unskip~\citet{314005:7060336}.

\textbf{Potential approach:} Since the WCP is prone to a diverse range of malicious activities, an intelligent technique is required to tackle this challenge. Although the vulnerabilities of distributed processing, software, and wireless channels threaten the WCP, they could be clues for developing a robust security scheme for the WCP, owing to BlockChain (BC)\unskip~\citet{314005:7023681} technology, and the moving target defense (MTD)\unskip~\citet{314005:7023683}.  The former could utilize the distributed entities (controllers) as trustworthy distributed ledgers to verify the integrity of the exchanged information among distributed controllers. It involves two powerful techniques: proof of work (POW) by miners (distributed entities) to attain the security requirements and chain of cryptographic blocks to guarantee the integrity of the exchanged information. It can be simply defined as a linked list (chain) data structure, where each element in the list is a cryptographic block. Each block in the chain has patches of an encrypted valid process (transaction) in addition to an encoded hash of the previous block in the chain. Thus, a chain of blocks is constructed in an iterative manner. The blockchain could also be seen as a distributed database shared among all distributed participants. Thus, an attacker needs to attack all nodes in the chain to collude against the network. In the latter, MTD provides a range of uncertainties in system configurations to complicate the system for attackers. The WCP's programmability features endow a myriad of system configurations because it deploys SDR and SDN to represent all layers of the protocol stack. MTD leverages this ability to randomly alter their configuration. \mbox{}\protect\newline To exemplify, recall the security API in the D-WCP architecture, which is shown in Figure~\ref{figure-d367bdee6cfddf21405f3e05ae6e0bb5}. The diversity in the PBS and the wireless controller's SDR interface configurations would be achieved by incorporating MTD and the adaptive modulation and coding (AMC) information base (AMCIB) into the PBS's security layer and wireless controller's security API, respectively. The AMCIB provides a diverse range of radio parameters that would be utilized by MTD to reconfigure the modulation and coding schemes of the radio signals. These MTD-based radio reconfigurations would secure the radio signal that will be transmitted over the wireless link as well as the PBS. The diversity in the SDN controller system configurations would be achieved by incorporating a virtualization layer with the control logic layer to virtualize the controller's resources, and hence, a diverse range of the controller system configurations would be available for the MTD-based security agent. Thus, the controller's MTD-based security agent would provide a two-dimensional controller and radio configuration security scheme.\mbox{}\protect\newline Nevertheless, both BC and MTD have inherent limitations regarding their need for high processing power to process their algorithms. Thus, a smart algorithmic approach is still needed to address this challenge\unskip~\citet{314005:7023679} for the WCP design.

\subsection{How would SDN and SDR be synergistic for 5G and Beyond?}\textbf{Challenge:} To provide a robust programmable framework for the WCP, SDN and SDR should cooperate harmonically. In fact, we are still far from this harmonic inter-operation of SDN with SDR because there is still the challenge of how the SDN controller directly programs the SDR-based PBS rather than only controlling its behavior\unskip~\citet{314005:7023678}. Although they share the concept of abstracting hardware components and representing their functionalities with flexible software, the differential nature of the data that each software paradigm needs to process limits the achievement of real harmonization between the SDR and SDN. To illustrate, SDR software manipulates data, which are stochastic signals that are handled over stochastic systems (wireless channels) that mandate running sophisticated signal processing functions, such as fast Fourier transform (FFT). On the other hand, SDN software manipulates deterministic data that might have certain stochastic properties (e.g., control packets that have stochastic arrival times). Therefore, the challenge is how to converge these two software paradigms, which have to manipulate data with different characteristics. In other words, the challenge is how to achieve a robust convergence between Pyretic and GNU radio.

\textbf{Potential approach:} Although there are noticeable efforts to integrate SDN with SDR for 5G and beyond, such as\unskip~\citet{314005:7023678,314005:7023632}, they have tackled the problem from only the reconfiguration and virtualization perspectives\unskip~\citet{314005:7023678}\unskip~\citet{314005:7023675}. In the proposed D-WCP architecture, the wireless controller could achieve this convergence through its softradio API, as shown in Figure~\ref{figure-d367bdee6cfddf21405f3e05ae6e0bb5}. It is envisaged that the softradio API could work as a shared module between two different software models (GNU radio and SDN control logic). To exemplify, consider the deployment of shared memory to manage the operation between two different software, such as ns3 and Matlab\unskip~\citet{314005:7023680}. However, this analogy is partially correct because the integration process is still carried out on software that handles deterministic data. Thus, the softradio API could not only be a shared memory between two software paradigms but also could be a powerful interface that computationally models the stochasticity of signals and reforms it, using very fast algorithms, to data shapes that can be manipulated by Pyretic or any developed language that could address the challenge of integrating the programmability of SDN with SDR within a unified development platform.

\subsection{Stochastic optimization for DRL}\textbf{Challenge:} As discussed in Section 4, DRL constitutes different computational methods (e.g., the Markovian decision process, MDP, temporal difference, TD, hard attention and stochastic value gradient\unskip~\citet{314005:7022311,314005:7022493}) to optimally estimate the stochastic variable \textit{Q} and optimize the stochastic policy$\pi $such that the maximum long-term return is achieved. Since these optimizations problems are NP-hard, the topic of developing effective approximation algorithms is open for further research to obtain the best accurate approximation algorithm.

\textbf{Potential approach:} A stochastic optimization tool, which has successfully addressed the uncertainty in the wireless environment (e.g.,\unskip~\citet{314005:7023495,314005:7023499,314005:7023428,314005:7023584,314005:7023501}), has recently been deployed to provide an end-to-end learning scheme for different complex control systems\unskip~\citet{314005:7023682}. In this sense, it could be examined for DRL.\mbox{}\protect\newline Stochastic programming provides a powerful mathematical tool to handle optimization under uncertainty. It has been recently exploited to optimize resource allocation in various types of wireless networks operating under uncertainties (examples include\unskip~\citet{314005:7023496,314005:7023497,314005:7023498}). \mbox{}\protect\newline Stochastic optimization models the system's uncertainties as follows:

\begin{itemize}
  \item \relax By relaxing the stochastic design constraints so that they are only satisfied with a given probability. 
  \item \relax By modeling as many operational scenarios as possible, where each scenario occurs with a specific probability. 
\end{itemize}
  As DRL utilizes the real trajectories to form the controller's database, these trajectories could provide the scenario space for the stochastic optimization to accurately estimate the occurrence probability of each scenario and, thus, provide another estimation tool to optimize\textit{ Q} and$\pi $. 

Table~\ref{table-wrap-0de9d62519b131ae1a55bcc051c41618} summarizes the directions for future research.

\begin{table*}[t!]
\caption{{ Directions for future research} }
\label{table-wrap-0de9d62519b131ae1a55bcc051c41618}
\def\arraystretch{1}
\ignorespaces 
\centering 
\begin{tabulary}{\linewidth}{p{\dimexpr.33\linewidth-2\tabcolsep}p{\dimexpr.33\linewidth-2\tabcolsep}p{\dimexpr.34\linewidth-2\tabcolsep}}
\hline 
Topic & Major challenge & Potential approach\\
\tblmidrule 
D-WCP or I-WCP &
  Thorough quantitative comparison is required rather than relying on qualitative evaluation &
  Addressing the effect of interference in D-WCP when WDP placement and WCPP would be jointly optimized\\
WCP security &
  Distributed operation, relying on susceptible software, and vulnerable wireless channels raise the possibilities of attacking the WCP &
  Deploying Blockchain\unskip~\citet{314005:7023681}  and MTD\unskip~\citet{314005:7023683}\\
SDR with SDN &
  SDR software manipulates stochastic signals, but SDN processes deterministic flows with stochastic properties &
  A computational mapping algorithm could be deployed for stochastic signals to reform them to be deterministically implemented\\
Stochastic optimization for DRL &
  DRL stochastic estimations require an accurate approximation &
  Examining stochastic optimization for DRL\unskip~\citet{314005:7023682}\\
\tblbottomrule 
\end{tabulary}\par 
\end{table*}
Obviously, the road to WCP has serious obstacles, which need smart maneuvers to overcome them. However, it is envisaged that the WCP would be ready for 5G networks and beyond by unifying the current research efforts in stochastic optimization, stochastic geometry, deep reinforcement learning, digital signal processing, SDR, SDN, network programming languages, and network operating systems.
    
\section{Conclusions}
The proposed review of the recent 5G architectures indicates that software-based solutions (e.g., SDN and SDR) are significant components of 5G networks and beyond. To simplify the management of these networks, the SDN controller has to communicate with a massive number of SDR-based wireless data plane (WDP) devices. Two different controller-WDP communication approaches have been deployed in recent 5G systems: an indirect wireless connection via a master WDP element (MWDP) and a direct wireless connection between a wireless controller and WDP elements. Both WCP schemes, the direct wireless connection (D-WCP) and the indirect one (I-WCP), suffer from the challenge of transmitting critical control information over wireless channels where a diverse range of uncertainties are present. Accordingly, a robust WCP framework is needed to address this challenge.

Although these schemes have been proposed in the literature, the discussion of their design principles and challenges has been overlooked. Consequently, this paper has taken a step to provide an overview of the WCP design by performing a qualitative comparison between the two WCP schemes and using deep reinforcement learning (DRL) techniques to tackle the challenge of WCP design within an environment that is replete with uncertainties. 

In light of the conducted work, there are two major implications:

\begin{itemize}
  \item \relax SDN and 5G mutual benefit, in the context of the WCP design, SDN also needs 5G as 5G needs SDN. In I-WCP, for example, SDN controllers would rely on the 5G wireless resources to manage the underlying wireless elements. Similarly, in D-WCP, wireless SDN controllers would rely on the placement strategy of the 5G base stations to jointly optimize their placement.
  \item \relax The wireless controller's programmability, that in addition to what is mentioned in Section 5 about converging the programmability features of SDN and SDR, the DRL is also programmable and could contribute to the convergence of SDN and SDR under the umbrella of wireless controller's design.
\end{itemize}
  This paper has proposed a generic framework for the WCP design, which has been incorporated into a myriad of 5G networks and beyond using two schemes (direct and indirect), by using DRL principles to address the uncertainties in its design. However, it lacks a practical implementation of the proposed framework to quantitatively evaluate the performance of each WCP scheme.

To provide a robust framework for the WCP design, the following question needs to be answered: Is WCP ready for 5G networks and beyond? To provide an answer to this broad but interesting question, thorough research efforts are still required to address the following:

\begin{itemize}
  \item \relax Which is better, D-WCP or I-WCP? 
  \item \relax How can the integrity of the sensitive control information be guaranteed? 
  \item \relax How can SDN and SDR be synergistic for 5G and beyond?
  \item \relax Is stochastic optimization promising for DRL?
\end{itemize}
  WCP could be ready for 5G and beyond by synthesizing a complimentary vision from data science, stochastic geometry, stochastic optimization, SDN, software systems, digital signal processing, and SDR.

\section*{Acknowledgments}This research work was funded by the Center for Sustainable Mobility (CSM) at the Virginia Tech Transportation Institute (VTTI) under the sponsorship of the University Mobility and Equity Center, Blacksburg, Virginia, USA in collaboration with the ministry of higher education (MoHE), Egypt.

\section*{References}

\bibliographystyle{model2-names}

\bibliography{\jobname}

\section*{Author biography}\noindent
\textbf{EmadelDin A. Mazied} is currently a visiting scholar in the Bradley Department of Electrical Engineering at Virginia Tech. He has been a Ph.D. student in Electrical Engineering, at Alexandria University, Egypt. He obtained his B.Sc. of in Electronics Engineering from Menoufia University, Egypt. He began his research in VoIP over wireless networks in 2010 at Alexandria University. He received the M.Sc. in Electrical Engineering from Alexandria University in June 2012. In July 2013, he received a research assistantship in the networking and distributed Systems department in the City for Scientific Research. In August 2014, he received a scholarship for his Ph.D. studies at Alexandria University, including research at Virginia Tech. In October 2014, he received a teaching assistantship in the Electrical Engineering department in Sohag University, Egypt. His research interests include wireless communication networks, QoS in future generation wireless networks, and SDN for wireless networks.

\smallskip\noindent 
\textbf{Mustafa Y. ElNainay} is an Associate Professor of the Computer and Systems Engineering department at Alexandria University, Egypt. He is also the associate director of the Virginia Tech-Middle East and North Africa (VT-MENA) program for administration and research and adjunct faculty at Virginia Tech. He received his B.Sc. and M.Sc. in Computer Science from Alexandria University in 2001 and 2005 respectively and his Ph.D. in Computer Engineering from Virginia Tech in 2009. His research interests include wireless and mobile networks, cognitive radio and cognitive networks, and software testing automation and optimization. He is the receipt of ICDT best paper award. He is a member of IEEE and served as a reviewer for various international conferences.

\smallskip\noindent 
\textbf{Mohammad J. Abdel-Rahman} is currently an Assistant Professor in the Electrical Engineering Department at Al Hussein Technical University, Amman, Jordan. He is also an adjunct professor in the Bradley Department of Electrical and Computer Engineering (ECE) at Virginia Polytechnic Institute and State University (Virginia Tech), USA. He was appointed a Research Associate position in the ECE Department at Virginia Tech, from January 2015 to August 2017. He received his PhD degree from the Electrical and Computer Engineering Department at the University of Arizona in 2014. He is the recipient of the College of Engineering 2014 Outstanding Graduate Student Award. He received his MSc degree from the Electrical Engineering Department at Jordan University of Science and Technology, Jordan, in 2010, and his BSc degree from the Communication Engineering Department at Yarmouk University, Jordan, in 2008. Dr. Abdel-Rahman's research is in the broad area of wireless communications and networking, with particular emphasis on resource management, adaptive protocols, and security issues. He serves as a reviewer for several international conferences and journals. He is a member of the IEEE.

\smallskip\noindent 
\textbf{Scott F. Midkiff} is currently a professor in the Bradley Department of Electrical and Computer Engineering at Virginia Polytechnic Institute and State University (Virginia Tech), and the vice president for information technology and chief information officer at Virginia Tech, Blacksburg, VA, USA. From 2009 to 2012, he was the department head of the Bradley Department of Electrical and Computer Engineering at Virginia Tech. From 2006 to 2009, he served as a program director at the US National Science Foundation. His research interests include wireless and ad hoc networks, network services for pervasive computing, and cyber-physical systems. He is a senior member of the IEEE.

\smallskip\noindent 
\textbf{Mohamed R. M. Rizk} obtained his B.Sc. from Alexandria University and his master's and Ph.D. from McMaster University, Canada. He worked as an assistant professor at McMaster University. He was a visiting professor at Sultan Qaboos university, Oman, Beirut Arab University and the Arab Academy for Science and Technology, Egypt. He is an Adjunct professor to Virginia Polytechnic and State University, Virginia, U.S.A.  He is a Life Senior of the IEEE and the Chairman of the IEEE Alexandria subsection. He co-authored 110 conference and journal papers, and co-authored the book ``Medium Access Control for Multimedia Wireless Systems'' and the book chapter ``Advanced Methods for Complex Network Analysis''. His research interests include Computer Aided Design, Wireless networks, Encryption, Fuzzy Logic, Image processing, cognitive radio and Nano Technology.

\smallskip\noindent 
\textbf{Hesham A. Rakha} received his B.Sc. degree (with honors) in Civil Engineering from Cairo University, Cairo, Egypt, in 1987 and M.Sc. and Ph.D. degrees in Civil and Environmental Engineering from Queen's University, Kingston, ON, Canada, in 1990 and 1993, respectively. He is currently the Samuel Reynolds Pritchard Professor of Engineering in the Charles E. Via, Jr. Department of Civil and Environmental Engineering, a Courtesy Professor in the Bradley Department of Electrical and Computer Engineering and the Director of the Center for Sustainable Mobility (CSM) at the Virginia Tech Transportation Institute (VTTI). He is a Professional Engineer in Ontario and a member of the Institute of Transportation Engineers (ITE), the American Society of Civil Engineers (ASCE), the Institute of Electrical and Electronics Engineers (IEEE), the Society of Automotive Engineers (SAE), and the Transportation Research Board (TRB). He is on the Editorial Board of the Transportation Letters, IET Intelligent Transport Systems Journal, and the International Journal of Transportation Science and Technology. In addition, he is an Associate Editor for the IEEE Transactions of Intelligent Transportation Systems and the Journal of Intelligent Transportation Systems. Dr. Rakha's areas of research include traffic flow theory, traveler and driver behavior modeling, dynamic traffic assignment, transportation network control, use of artificial intelligence in transportation, intelligent vehicle systems, connected and automated vehicles, transportation energy and environmental modeling, and transportation safety modeling.

\smallskip\noindent 
\textbf{Allen B. MacKenzie} is an Associate Professor in the Bradley Department of Electrical and Computer Engineering at Virginia Tech, where he has been on the faculty since 2003. He is the associate director of Wireless @ Virginia Tech. During the 2012-2013 academic year, he was an E. T. S. Walton Visiting Professor at Trinity College Dublin. Prof. MacKenzie's research focuses on wireless communications systems and networks. His current research interests include cognitive radio and cognitive network algorithms, architectures, and protocols and the analysis of such systems and networks using game theory and stochastic optimization. His past and current research sponsors include the National Science Foundation, Science Foundation Ireland, the Defense Advanced Research Projects Agency, and the National Institute of Justice. He is a senior member of the IEEE and a member of the ASEE and the ACM. Prof. He is an area editor of the "IEEE Transactions on Communications" and an associate editor of the "IEEE Transactions on Cognitive Communications and Networking". He is the author of more than 90 refereed conference and journal papers and a co-author of the book Game "Theory for Wireless Engineers".

\end{document}